**Elastic theory of ideal liquid crystal elastomers**

Zumrat Usmanova, Ruobing Bai*

Department of Mechanical and Industrial Engineering, College of Engineering, Northeastern University, Boston, MA 02115, USA

*Corresponding author: ru.bai@northeastern.edu

**Abstract**

A liquid crystal elastomer (LCE) exhibits a two-way coupling between its mesogen order and network elasticity: the mesogen order sets the undeformed network conformation, and the network deformation reorients the mesogens. This coupling gives rise to large, reversible stimuli-responsive actuation, as well as to soft elasticity, a finite deformation at nearly zero stress. This article reviews the elastic theory of ideal LCEs, bridging single-chain statistics and continuum constitutive models. The "ideal" elastic behavior rests on the assumption of anisotropic Gaussian chains that retain no memory of the crosslinking state, such that rotating the director costs no free energy. The corresponding neo-classical free energy of a monodomain LCE is non-quasi-convex, with a continuous family of stress-free ground states, leading to thermally induced spontaneous deformation and soft elasticity accommodated by coexisting stripe domains. The quasi-convex relaxation of the free energy classifies the behavior of polydomain LCEs into three phases, each with a distinct macroscopic mechanical response depending on the stretch state. The ideal elastic theory can be further extended to incorporate additional behaviors and physical processes, including thermo- and photomechanical coupling, Frank elasticity from director distortion, semi-soft or non-soft behavior arising from network nonideality, and temperature- and rate-dependent viscoelasticity.

## 1. Introduction

Liquid crystal elastomers (LCEs) are rubbery polymer networks that incorporate rod-like liquid crystal molecules, called *mesogens*, into their chains. The mesogens retain enough mobility to order or disorder orientationally like a liquid crystal, while the crosslinked polymer network holds the chains and mesogens together like a rubber. This synergy produces a two-way coupling between the mesogens and their attached polymer chains: when the mesogens order along a common direction, known as the *director*, the chains deform; conversely, when a chain is stretched by an external load, the mesogens rotate along with the stretch. LCEs are further classified as either *side-chain* or *main-chain* depending on how mesogens are linked to the network. In side-chain LCEs, mesogens hang from the backbone as pendant groups and therefore couple only weakly to chain conformation. In main-chain LCEs, mesogens are built into the backbone itself, so their ordering deforms the chain directly. Because of this stronger coupling, main-chain LCEs dominate the recent literature and will be the focus of this review.

Most LCEs are thermotropic with the phase of their mesogens set by temperature. Above a nematic-isotropic transition temperature $T_{\mathrm{ni}}$, typically around 60-70 °C in modern thiol-acrylate LCEs, the mesogens point in random directions (*isotropic* phase), and the elastomer is isotropic like an ordinary rubber. Below $T_{\mathrm{ni}}$, the mesogens develop long-range orientational order along the director (*nematic* phase), and their attached chains elongate in the same direction. Heating a nematic LCE with uniformly aligned mesogens (i.e., *monodomain*) to temperatures above $T_{\mathrm{ni}}$ induces uniaxial contraction by tens of percent, and cooling restores the original shape. This spontaneous and reversible two-way thermomechanical actuation is the most studied stimuli-responsive behavior of LCEs and underlies their applications in actuation and shape morphing.

The existence of LCEs and their thermomechanical actuation were first predicted theoretically by de Gennes in 1975[1], and the first such material was synthesized by Finkelmann et al. in 1981 by lightly crosslinking side-chain liquid crystalline polysiloxanes[2]. Küpfer and Finkelmann further achieved a uniform director in experiment through two-stage crosslinking, with a mechanical pre-stretch in the second stage[3]. Theoretically, Bladon, Terentjev, and Warner developed the pioneering neo-classical model for LCE elasticity by extending the Gaussian chain model of rubber elasticity to anisotropic chains[4]. The theory successfully predicted soft elasticity in LCEs, a finite deformation at almost no stress[5, 6]. Soft elasticity was subsequently validated in experiments by Küpfer and Finkelmann[7], together with the formation of microscopic stripe domains during deformation observed by Kundler and Finkelmann[8]. The theoretical model has since been refined and extended substantially, including the addition of nonideality by Verwey and Warner[9, 10], the quasi-convex relaxation of the neo-classical energy for polydomain LCEs by Conti, DeSimone, and Dolzmann[11-14], and the effect of synthesis and processing on soft elasticity by Biggins, Warner, and Bhattacharya[15, 16]. In parallel, Finkelmann et al. showed in experiment that an LCE can also be actuated by light through photochemistry[17]: illuminating an LCE containing azobenzene isomerizes the azobenzene from a rod-like to a bent shape, dilutes the nematic order, and contracts the elastomer. In 2003, Warner and Terentjev provided a comprehensive review of the theory and experiment of LCEs in their seminal book[18].

A more recent wave of development over the past decade has been largely enabled by advances in LCE chemistry. Both the thermo- and photo-responsive actuation of LCEs require mesogens to be aligned along a uniform director locally or globally, which is typically achieved by introducing an additional external field before, during, or after crosslinking. In 2015, Yakacki, Saed, and coworkers introduced a two-stage fabrication method analogous to that of Küpfer and Finkelmann, requiring only ordinary laboratory equipment[19-21]: a thiol-acrylate reaction forms the network in the first stage, and free-radical polymerization under mechanical pre-stretch aligns the mesogens in the second. Besides mechanical pre-stretch, mesogen alignment is also routinely achieved by rubbed or patterned substrate surfaces, by magnetic fields, and by photo-alignment, which writes the director in local, microscale domains such that more complex actuation and shape morphing are achieved by heterogeneous director fields[22-24]. These advances in chemistry have in turn enabled new theories, experiments, and applications. To name a few, the elastic response has been studied under complex mechanical loads beyond uniaxial tension, such as equal- and non-equal-biaxial loading[25]; the roles of genesis and pre-stretch in thermomechanical properties have been quantified and modeled[26, 27]; cyclic and rate-dependent experimental, theoretical, and numerical studies have emerged[28-34]; thermo- and photo-responsive actuation of LCEs have been built into soft robots, shape-morphing sheets, and biomedical devices[22, 23, 35-51]; and highly dissipative LCEs have been used for damping and impact protection[52-54], energy-absorbing architected structures[55-59], and adhesives[60-67].

This review is motivated by two practical needs. First, we intend this article as a self-contained entry point for researchers new to LCE mechanics. Assuming only a first graduate course in nonlinear elasticity, it connects the statistical mechanics of polymer chains to constitutive models in the language of continuum mechanics, and incorporates the major mathematical results from the calculus of variations. Second, because the field has grown at the intersection of several communities, the notation and vocabulary used across the literature have diverged and need reconciliation. For example, the tensor that represents the chain anisotropy in LCEs goes by several names and different normalized forms, and different models adopt different reference states, sometimes even within the same model across different loading conditions. Consequently, free energies that look different are often essentially identical, and free energies that look alike are sometimes not. We aim to resolve these ambiguities by deriving the theories systematically under a unified terminology. For a more comprehensive discussion of the physical foundation of the theories presented here, we refer the reader to the classic book of Warner and Terentjev[18], along with recent reviews of other aspects of LCEs[24, 68-75].

The paper is organized as follows. We begin by introducing thermotropic liquid crystals and the nematic order parameter (Section 2). We then build the neo-classical theory of monodomain LCEs from single-chain statistics, and use it to explain spontaneous deformation, stress-free states, stripe domains, and soft elasticity (Section 3). We extend the theory to polydomain LCEs as the quasi-convex relaxation of the same free energy, and work out the uniaxial, equal-biaxial, and non-equal-biaxial responses (Section 4). We further discuss four additional factors left out by the ideal theory: thermomechanical and photomechanical coupling (Section 5.1), Frank elasticity (Section 5.2), network nonideality (Section 5.3), and viscoelasticity (Section 5.4). We summarize our discussions in Section 6.

**2. Thermotropic liquid crystals**

Most LCEs of practical interest are thermotropic: the phase state of their LC mesogens is controlled by temperature (Fig. 1a). A nematic-isotropic transition temperature $T_{\mathrm{ni}}$ is accordingly defined. Above $T_{\mathrm{ni}}$, the rod-like mesogens are orientationally disordered and are in the so-called *isotropic* phase. Below $T_{\mathrm{ni}}$, mesogens develop long-range orientational order and fluctuate about a common axis, described by a unit vector called the director $\mathbf{n}$, defining the *nematic* phase. The nematic phase has no long-range positional order of the mesogen centers of mass.

In the nematic phase, the long-range orientational order about the director $\mathbf{n}$ is measured by a scalar *order parameter*

$$Q = \langle \frac{3}{2}\cos^2\theta - \frac{1}{2} \rangle, \tag{1}$$

where $\langle\ \rangle$ denotes the average over the orientational distribution of all mesogens, and $\theta$ is the angle between an individual mesogen and the director $\mathbf{n}$. The choice of the second Legendre polynomial, $(3\cos^2\theta - 1)/2$, is dictated by symmetry: since a nematic is quadrupolar rather than polar, i.e., the two ends of the director are indistinguishable ($\mathbf{n} \leftrightarrow -\mathbf{n}$), the lowest non-trivial measure of alignment is the second moment. Consequently, $Q$ measures the degree of alignment along the director $\mathbf{n}$: $Q = 1$ corresponds to perfect alignment ($\theta = 0$), $Q = 0$ to a fully isotropic phase, and $0 < Q < 1$ to the general nematic phase. Furthermore, the sign of $Q$ is physically meaningful: $Q > 0$ describes the usual prolate, rod-like alignment along $\mathbf{n}$, whereas $Q < 0$ describes oblate ordering, with $Q = -1/2$ corresponding to mesogens fully confined to the plane perpendicular to $\mathbf{n}$. Oblate LCEs are far less explored than the prolate ones with $Q > 0$ that dominate the literature, and are therefore beyond the scope of this review. Equivalently, $Q$ and $\mathbf{n}$ can be combined to form a symmetric order-parameter tensor $Q_{ij} = Q\left(3n_in_j - \delta_{ij}\right)/2$, where $Q$ is the scalar amplitude and $\mathbf{n}$ is the principal axis of the tensor.

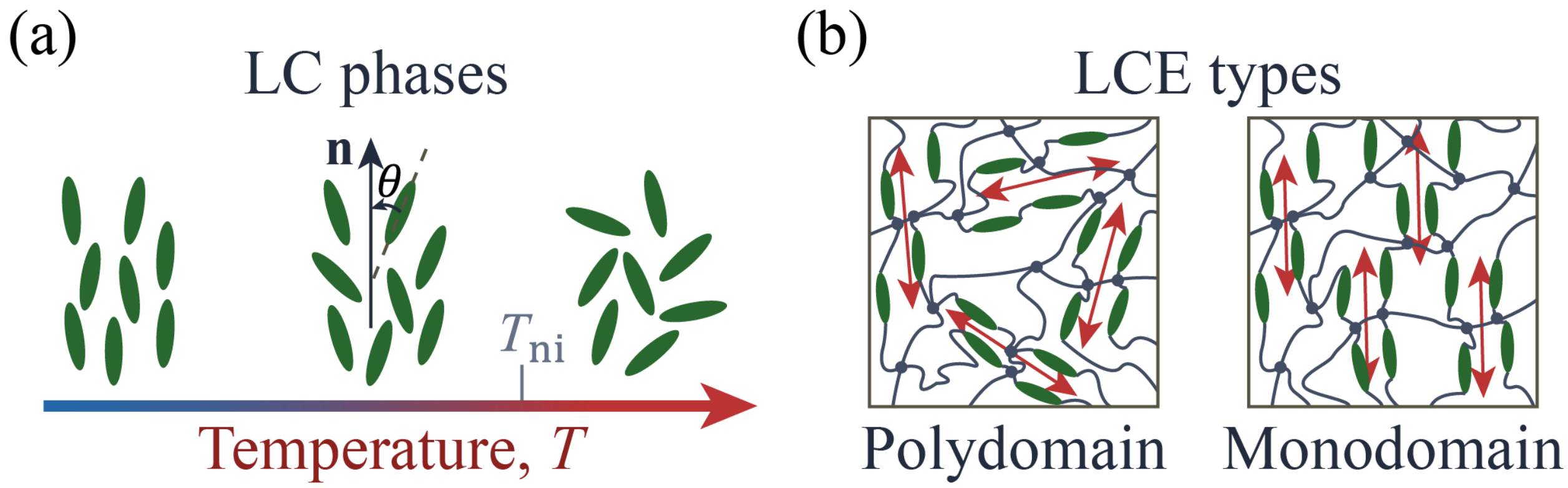

Fig. 1. Phases and types of thermotropic LCs and LCEs. (a) The phase state of thermotropic LC mesogens is controlled by temperature, $T$. Below the nematic-isotropic transition temperature $T_{\mathrm{ni}}$, mesogens (green rods) develop long-range orientational order and fluctuate about a common axis, described by a unit vector called the director $\mathbf{n}$, defining the *nematic* phase. Above $T_{\mathrm{ni}}$, mesogens are orientationally disordered and are in the so-called *isotropic* phase. (b) In the nematic phase of an LCE, the director $\mathbf{n}$ (indicated by red arrows) may vary from region to region (*polydomain* LCE, left), or stay uniform throughout the material (*monodomain* LCE, right).

The nematic-isotropic phase transition of thermotropic LCs can be described by the temperature-dependent order parameter $Q(T)$. At temperature $T < T_{\mathrm{ni}}$ in the nematic phase, $0 < Q(T) < 1$ and decreases with $T$. At $T \geq T_{\mathrm{ni}}$, $Q(T) = 0$ represents the isotropic phase. Most common LCEs have $Q \approx$ 0.5-0.6 at room temperature[18, 26]. We will further discuss the theory of this nematic-isotropic phase transition, its temperature-dependent $Q(T)$, and the corresponding thermomechanical coupling in LCEs in Section 5.1.

To form an LCE, mesogens are further linked into a rubbery polymer network via covalent bonds, either along the polymer backbone (*main-chain LCEs*) or as pendant side groups (*side-chain LCEs*). As discussed in Section 1, main-chain LCEs typically provide a stronger coupling between mesogens and polymer chains and are thus the focus of most modern studies including the current paper. Furthermore, as illustrated in Fig. 1b, in the nematic phase, the director $\mathbf{n}$ may vary from region to region (*polydomain* LCE), or stay uniform throughout the material (*monodomain* LCE), analogous to polycrystalline and single-crystal metals. The elastic theory and behavior of each type will be discussed in Sections 3 and 4, respectively.

As will be discussed in Section 5.1, the detailed temperature dependence of $Q(T)$ can be modeled by a free energy of mesogen interaction, such as the phenomenological Landau–de Gennes polynomial[1, 18] or the molecular Maier-Saupe model[76, 77]. Theoretical studies further suggest that mechanical stretch affects $Q$ only within a narrow temperature window around $T_{\mathrm{ni}}$[78-80]; outside this window, $Q$ is set almost entirely by temperature. Therefore, in Sections 3 and 4, we adopt the standard convention in LCE modeling to treat $Q$ as a function of temperature alone, $Q(T)$, independent of mechanical stretch.

**3. Monodomain LCEs**

A monodomain LCE has a uniform director $\mathbf{n}$ in the nematic phase. Because the mesogens are linked to the polymer network, their orientational order is transmitted to the polymer chains (Fig. 2): as mesogens align along the director $\mathbf{n}$, the polymer chain conformation becomes anisotropic, elongating along $\mathbf{n}$ and contracting in the transverse directions. When heated above $T_{\mathrm{ni}}$, the orientational order vanishes, and the conformation of polymer chains becomes isotropic like a common rubber. This intrinsic coupling between mesogen alignment and chain conformation gives monodomain LCEs their defining thermomechanical signature: an anisotropic elastic response in the nematic phase ($T < T_{\mathrm{ni}}$), large, reversible spontaneous deformation upon heating and cooling, and isotropic common-rubber-like behavior in the isotropic phase ($T > T_{\mathrm{ni}}$). The monodomain theory is also the foundation for polydomain LCEs, since the latter can be regarded as a collection of geometrically compatible monodomains with sizes as small as microscale (Section 4).

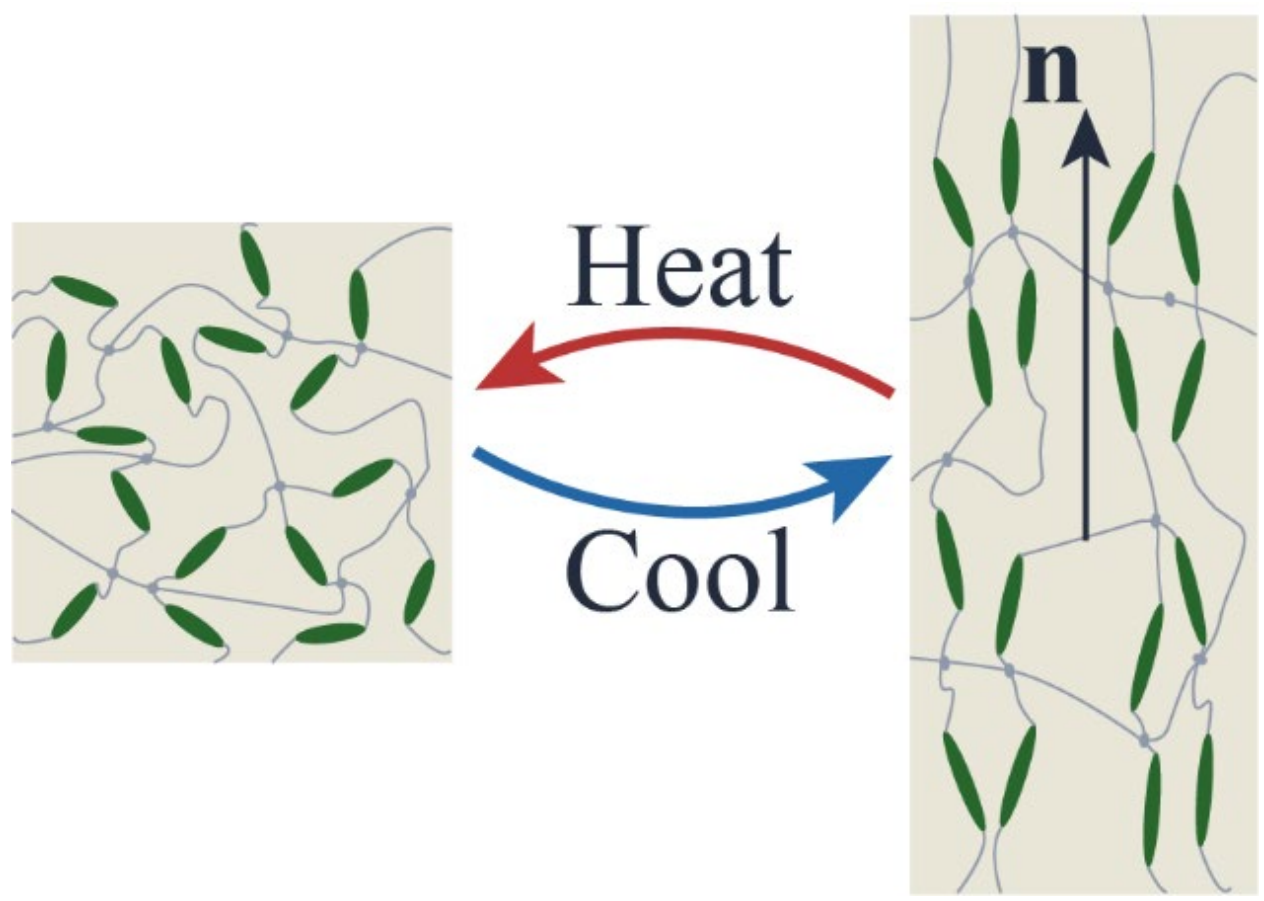

Fig. 2. Thermal-induced spontaneous shape change and molecular process of a monodomain LCE. At a temperature below $T_{\mathrm{ni}}$ (nematic phase, right), a monodomain LCE has a uniform director $\mathbf{n}$. When heated above $T_{\mathrm{ni}}$ (isotropic phase, left), the orientational order vanishes, and the conformation of polymer chains becomes isotropic like a common rubber.

The statistical-to-continuum theory that captures these behaviors of monodomain LCEs was established by Bladon, Terentjev, and Warner[4, 18], who extended the classical Gaussian chain model of rubber elasticity to include the anisotropic distribution of conformation set by the nematic order. In this section, we reintroduce this derivation. We begin with the single-chain statistical mechanics that yields the anisotropic elastic energy (Section 3.1). From it, we derive the stress-free ground state and the corresponding spontaneous deformation (Section 3.2). We then extend the energy to the tensor form (Section 3.3), generalize the stress-free ground states (Section 3.4), and use these to explain stripe domains and soft elasticity (Section 3.5). The theory and its outcomes lead to the definition of ideal LCEs (Section 3.6). Finally, we introduce alternative forms of the free energy using a different reference state (Section 3.7) or normalized terms (Section 3.8).

### 3.1 Scalar formulation of the anisotropic Gaussian chain model

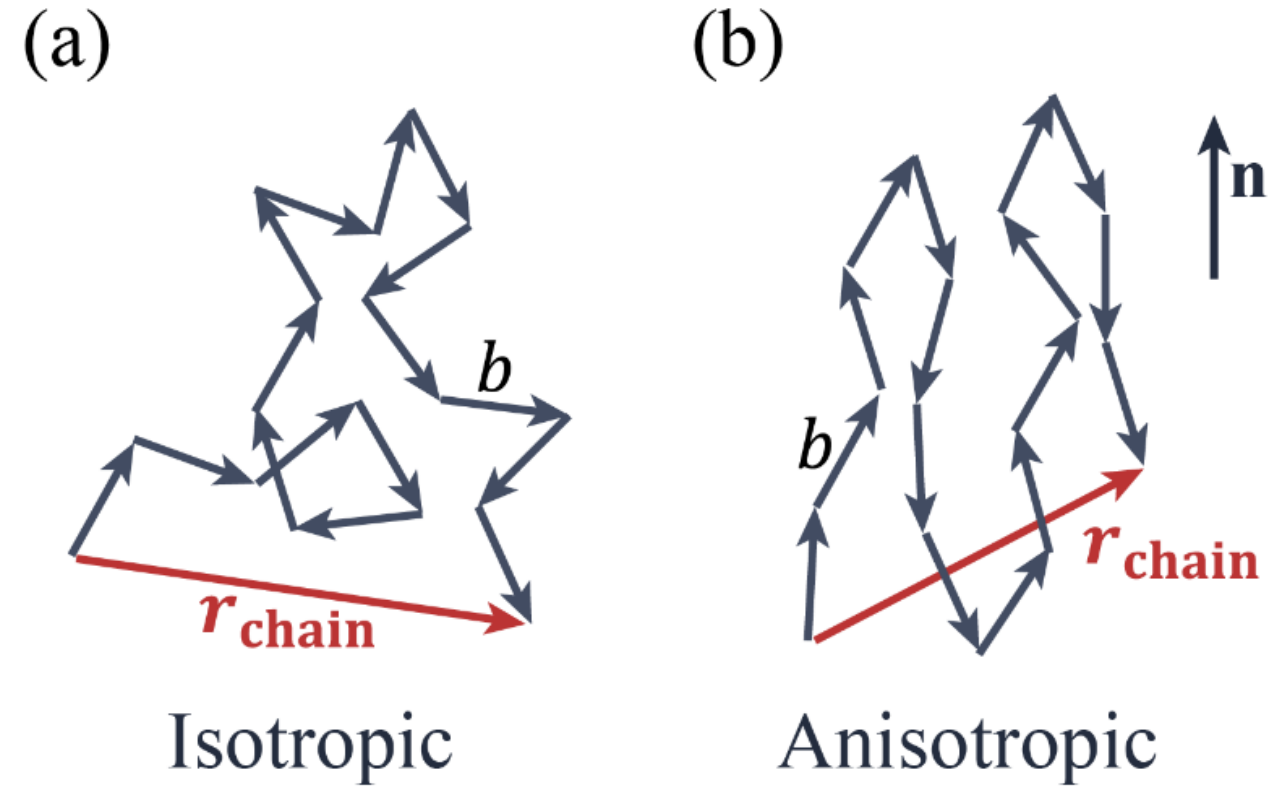

Fig. 3. Schematics of the isotropic and anisotropic freely jointed chain (Gaussian chain) model. (a) The classical model for an ordinary isotropic rubber treats the polymer chain as an isotropic random walk of $n_{\mathrm{chain}}$ steps with step length $b$. (b) The anisotropic model for a nematic LCE treats the polymer chain as an anisotropic random walk with segment orientations biased toward the director **n** (black arrow). The red arrow represents the end-to-end vector $\mathbf{r_{chain}}$.

Consider first the classical freely jointed chain (Gaussian chain) model for an ordinary isotropic rubber, which treats a polymer chain as a random walk of $n_{\mathrm{chain}}$ steps with step length $b$ (we use $n_{\mathrm{chain}}$ to differentiate it from the nematic director **n**). Each monomer step **u** is statistically isotropic with zero mean, $\langle u_x \rangle = \langle u_y \rangle = \langle u_z \rangle = 0$, and non-zero mean squares, $\langle u_x^2 \rangle = \langle u_y^2 \rangle = \langle u_z^2 \rangle = b^2/3$, where $u_x$, $u_y$, and $u_z$ are projections along the three axes. Summing $n_{\mathrm{chain}}$ independent steps and applying the central limit theorem, the end-to-end vector $\mathbf{r_{chain}}$ is Gaussian distributed (we use $\mathbf{r_{chain}}$ to differentiate it from the chain-anisotropy ratio $r$ defined later)[81], with the probability density function

$$p_{\mathrm{chain}}(\mathbf{r_{chain}}) = \left(\frac{3}{2\pi n_{\mathrm{chain}} b^2}\right)^{3/2} \exp\left(-\frac{3r_{\mathrm{chain}}^2}{2n_{\mathrm{chain}} b^2}\right). \tag{2}$$

The Helmholtz free energy of one chain is $w_e = -kT\log p_{\mathrm{chain}}$. This entropic spring underlies neo-Hookean elasticity.

Now consider a nematic LCE under the same freely jointed chain assumption, except that the orientation of each monomer step is biased by the mesogen orientational order. We set up a spherical coordinate system with the $z$ axis along the director **n**, such that each monomer forms a polar angle $\theta$ with the director. By symmetry, we have

$$\langle u_x \rangle = \langle u_y \rangle = \langle u_z \rangle = 0, \tag{3}$$

$$\langle u_x^2 \rangle = \langle u_y^2 \rangle = \langle (b\sin\theta\cos\varphi)^2 \rangle = \frac{1}{2} b^2 \langle \sin^2\theta \rangle = \frac{1}{3} b^2 (1 - Q), \tag{4}$$

and

$$\langle u_z^2 \rangle = b^2 \langle \cos^2\theta \rangle = \frac{1}{3} b^2 (1 + 2Q), \tag{5}$$

where $\varphi$ is the azimuthal angle. In Eqs. 4 and 5, we have used the definition of the nematic order parameter $Q$ in Eq. 1 to calculate the average.

Summing $n_{\text{chain}}$ independent steps and applying the central limit theorem, the end-to-end vector $\mathbf{r_{chain}}$ again follows a multivariate Gaussian distribution,

$$p_{\text{chain}}(\mathbf{r_{chain}}) = \left(\frac{3}{2\pi n_{\text{chain}} b^2}\right)^{3/2} \left(\frac{1}{(1-Q)^2(1+2Q)}\right)^{1/2} \exp\left[-\frac{3}{2n_{\text{chain}} b^2}\left(\frac{r_{\text{chain},x}^2}{1-Q} + \frac{r_{\text{chain},y}^2}{1-Q} + \frac{r_{\text{chain},z}^2}{1+2Q}\right)\right]. \quad (6)$$

The Helmholtz free energy of one nematic polymer chain is $w_e = -kT \log p_{\text{chain}}$, or

$$w_e = kT\left[\frac{3}{2n_{\text{chain}} b^2}\left(\frac{r_{\text{chain},x}^2}{1-Q} + \frac{r_{\text{chain},y}^2}{1-Q} + \frac{r_{\text{chain},z}^2}{1+2Q}\right) + \frac{1}{2}\log[(1-Q)^2(1+2Q)]\right], \quad (7)$$

omitting a constant.

Choosing the undeformed, stress-free elastomer in the isotropic phase ($T > T_{\text{ni}}$) as the reference state, the mean squares of end-to-end vector $\mathbf{R_{chain}}$ in the reference state along the three axes follow

$$\langle R_{\text{chain},x}^2 \rangle = \langle R_{\text{chain},y}^2 \rangle = \langle R_{\text{chain},z}^2 \rangle = \frac{1}{3} n_{\text{chain}} b^2. \quad (8)$$

Summing and averaging over $N$ chains per unit volume and assuming affine deformation, $r_{\text{chain},x} = \lambda_1 R_{\text{chain},x}$, $r_{\text{chain},y} = \lambda_2 R_{\text{chain},y}$, and $r_{\text{chain},z} = \lambda_3 R_{\text{chain},z}$, we obtain the Helmholtz free energy density of the nematic elastomer

$$W(\lambda_1, \lambda_2, \lambda_3; Q) = \frac{NkT}{2}\left[\frac{\lambda_1^2}{1-Q} + \frac{\lambda_2^2}{1-Q} + \frac{\lambda_3^2}{1+2Q} + \log[(1-Q)^2(1+2Q)]\right]. \quad (9)$$

Here, $\lambda_1$, $\lambda_2$, and $\lambda_3$ are the three principal stretches with $\lambda_3$ aligned along the director $\mathbf{n}$, satisfying material incompressibility

$$\lambda_1 \lambda_2 \lambda_3 = 1. \quad (10)$$

The logarithmic term reflects the change in the volume of the chain-conformation distribution from the reference (isotropic) to the current (nematic) state, independent of the deformation. In the isotropic phase, $Q = 0$, Eq. 9 reduces to the conventional isotropic incompressible neo-Hookean model.

### 3.2 Stress-free ground state

The Helmholtz free energy in Eq. 9 can be further written in its incompressible form by substituting $\lambda_2 = 1/(\lambda_1 \lambda_3)$ from Eq. 10,

$$W(\lambda_1, \lambda_3; Q) = \frac{NkT}{2}\left[\frac{\lambda_1^2}{1-Q} + \frac{1}{\lambda_1^2 \lambda_3^2 (1-Q)} + \frac{\lambda_3^2}{1+2Q} + \log[(1-Q)^2(1+2Q)]\right]. \quad (11)$$

The principal true (Cauchy) stresses thereby follow the relationship

$$\begin{cases} \sigma_1 - \sigma_2 = \lambda_1 \dfrac{\partial W(\lambda_1, \lambda_3; Q)}{\partial \lambda_1} \\ \sigma_3 - \sigma_2 = \lambda_3 \dfrac{\partial W(\lambda_1, \lambda_3; Q)}{\partial \lambda_3} \end{cases}. \tag{12}$$

When the elastomer is in a stress-free ground state, $\sigma_1 = \sigma_2 = \sigma_3 = 0$. Substituting this into Eq. 12, we obtain

$$\begin{cases} \lambda_3 = \lambda_\parallel = r^{1/3} = \left(\dfrac{1+2Q}{1-Q}\right)^{1/3} \\ \lambda_1 = \lambda_2 = \dfrac{1}{\sqrt{\lambda_\parallel}} = \lambda_\perp = r^{-1/6} = \left(\dfrac{1+2Q}{1-Q}\right)^{-1/6} \end{cases}, \tag{13}$$

where we have defined $\lambda_\parallel$ as the stress-free (ground-state) stretch along the nematic director, $\lambda_\perp$ as the stress-free stretch perpendicular to it, and $r = (1+2Q)/(1-Q)$ as the *chain-anisotropy ratio* in the LCE.

Eq. 13 reveals the following physical picture. Recall that we have chosen the undeformed, stress-free elastomer in the isotropic phase ($T > T_{\mathrm{ni}}$) as the reference state, where $Q = 0$, $r = 1$, and we obtain $\lambda_1 = \lambda_2 = \lambda_3 = 1$ from Eq. 13, as expected. By contrast, consider a current state in which the LCE is stress-free but in the nematic phase ($T < T_{\mathrm{ni}}$), such that $0 < Q < 1$, $r > 1$, and we obtain $\lambda_3 = \lambda_\parallel = r^{1/3} > 1$ and $\lambda_1 = \lambda_2 = \lambda_\perp = r^{-1/6} < 1$. That is, the monodomain nematic LCE exhibits a spontaneous (stress-free) deformation that elongates along the director $\mathbf{n} = \mathbf{e_3}$ while shrinking perpendicular to it (Fig. 4). Conversely, heating the nematic LCE causes it to contract along its director until returning to the isotropic phase at high temperature. Because the ideal model here assumes thermodynamic equilibrium, this thermally induced spontaneous deformation is reversible, which forms the basis for thermal actuation of nematic monodomain LCEs. At room temperature, common LCEs have $Q \approx 0.6$, such that $r \approx 5.5$ and $\lambda_\parallel \approx 1.8$. Equivalently, when heating such an LCE from room temperature to $T > T_{\mathrm{ni}}$, the spontaneous contractive stretch is about $\lambda_\parallel^{-1} = 0.6$.

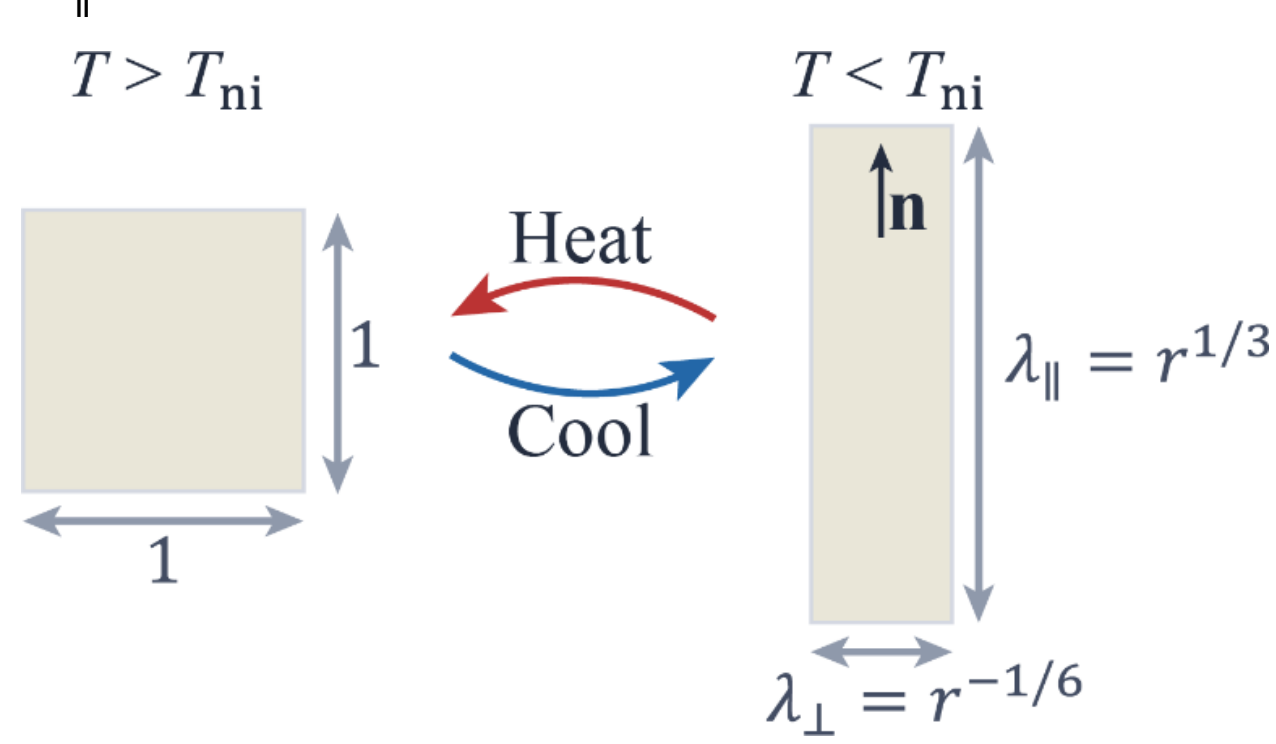

Fig. 4. Theoretically predicted thermal-induced spontaneous shape change of a monodomain LCE at the continuum scale. In the isotropic phase ($T > T_{\mathrm{ni}}$, $Q = 0$), the undeformed sample has $\lambda_1 = \lambda_2 = \lambda_3 = 1$. Cooling to the nematic phase ($T < T_{\mathrm{ni}}$) produces a stress-free spontaneous stretch along the director $\mathbf{n}$, $\lambda_\parallel = r^{1/3}$, and a corresponding contraction perpendicular to it, $\lambda_\perp = r^{-1/6}$, where $r = (1+2Q)/(1-Q)$ is the chain-anisotropy ratio. Heating back through $T_{\mathrm{ni}}$ reverses the deformation.

### 3.3 Continuum framework and tensor form

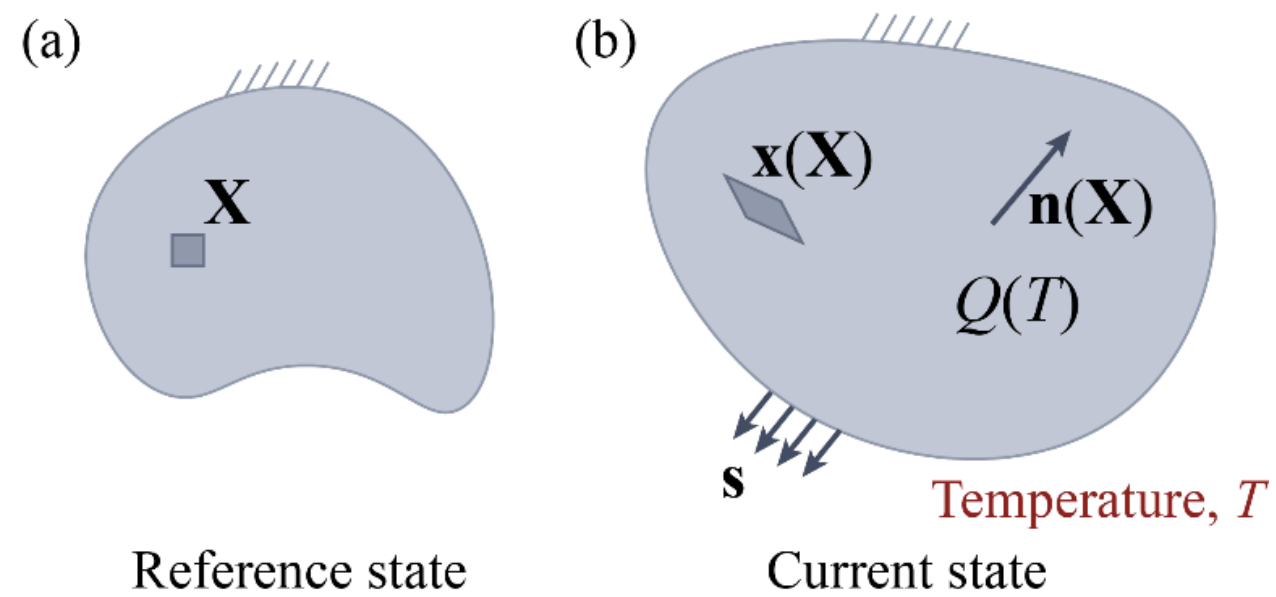

Fig. 5. Continuum framework for a nematic LCE. (a) The stress-free, undeformed isotropic phase is taken as the reference state, with a material particle labeled by its position **X**. (b) In the current state under a field of nominal stress **s**, the elastomer may stay at a different temperature $T$, have an order parameter $Q(T)$, deform to a shape defined by a field of current particle positions $\mathbf{x}(\mathbf{X})$, and bear a field of director $\mathbf{n}(\mathbf{X})$.

We next extend the theory to a general continuum framework (Fig. 5). Same as above, we take the stress-free, undeformed isotropic phase as the reference state, and label a material particle by its position **X**. Also in the reference state, the order parameter is zero and the director is indeterminate. In the current state, the elastomer may stay at a different temperature $T$, have an order parameter $Q(T)$, deform to a shape defined by a field of current particle positions $\mathbf{x}(\mathbf{X})$, and bear a field of director $\mathbf{n}(\mathbf{X})$. The deformation gradient is defined as

$$F_{iK} = \frac{\partial x_i}{\partial X_K}, \tag{14}$$

with material incompressibility

$$\det\mathbf{F} = 1. \tag{15}$$

Extending Eq. 6 to a nematic polymer chain with an arbitrary director **n**, the probability density function of the end-to-end vector $\mathbf{r}_{\mathbf{chain}}$ follows

$$p_{\text{chain}}(\mathbf{r}_{\mathbf{chain}}) = \left(\frac{3}{2\pi n_{\text{chain}} b^2}\right)^{3/2} \left(\frac{1}{\det\mathbf{l}}\right)^{1/2} \exp\left[-\frac{3}{2n_{\text{chain}} b^2}\left(r_{\text{chain},i} l_{ij}^{-1} r_{\text{chain},j}\right)\right], \tag{16}$$

where **l** is the *shape tensor* in the current state, expressed as

$$l_{ij} = (1 - Q)\delta_{ij} + 3Q n_i n_j, \tag{17}$$

or

$$\mathbf{l} = (1 - Q)\boldsymbol{\delta} + 3Q\mathbf{n} \otimes \mathbf{n}, \tag{18}$$

with its inverse

$$l_{ij}^{-1} = \frac{1}{1 - Q}\delta_{ij} + \left(\frac{1}{1 + 2Q} - \frac{1}{1 - Q}\right) n_i n_j, \tag{19}$$

or

$$\mathbf{l}^{-1} = \frac{1}{1-Q}\boldsymbol{\delta} + \left(\frac{1}{1+2Q} - \frac{1}{1-Q}\right)\mathbf{n}\otimes\mathbf{n}. \tag{20}$$

Further following the procedure in Eqs. 7-9 and taking $r_{\text{chain},i} = F_{iK}R_{\text{chain},K}$, we obtain the Helmholtz free energy density of the nematic elastomer in its tensor form

$$W(\mathbf{F},\mathbf{n};Q) = \frac{NkT}{2}\left[F_{iK}l_{ij}^{-1}F_{jK} + \log(\det\mathbf{l})\right], \tag{21}$$

or

$$W(\mathbf{F},\mathbf{n};Q) = \frac{NkT}{2}\left[\text{Tr}(\mathbf{F}^T\mathbf{l}^{-1}\mathbf{F}) + \log(\det\mathbf{l})\right]. \tag{22}$$

Eq. 22 is the *neo-classical model* first developed by Bladon, Terentjev, and Warner[4, 18]. In addition to the deformation gradient $\mathbf{F}$, the free energy of nematic elastomer has two more variables that describe its nematic state: the order parameter $Q(T)$ set by temperature and the unit director $\mathbf{n}(\mathbf{X})$ that serves as a kinematic internal variable, satisfying

$$n_i n_i = 1. \tag{23}$$

In equilibrium, the nominal (first Piola-Kirchhoff) stress tensor is calculated as

$$s_{iK}(\mathbf{F},\mathbf{n};Q) = -pH_{iK} + \frac{\partial W(\mathbf{F},\mathbf{n};Q)}{\partial F_{iK}}, \tag{24}$$

where $p$ is the Lagrangian multiplier that imposes incompressibility and $\mathbf{H} = \mathbf{F}^{-T}$. The equilibrium director field is governed by

$$\mathbf{n}\times\frac{\partial W(\mathbf{F},\mathbf{n};Q)}{\partial\mathbf{n}} = 0, \tag{25}$$

subject to unit norm (Eq. 23).

Diagonalizing $\mathbf{l}$ in Eq. 22 recovers the scalar form of free energy in Eq. 9. Further, in the isotropic phase ($T > T_{\text{ni}}$), $Q = 0$, $\mathbf{l} = \boldsymbol{\delta}$, and Eq. 22 reduces to the neo-Hookean energy $W = (NkT/2)\text{Tr}(\mathbf{F}^T\mathbf{F})$.

**3.4 General stress-free ground states**

Now we generalize the stress-free ground state from Section 3.2 by taking $s_{iK} = 0$ in Eq. 24. This is equivalent to assuming a homogeneous deformation and minimizing $W(\mathbf{F},\mathbf{n};Q)$ in Eq. 22 with respect to $\mathbf{F}$, subject to $\det\mathbf{F} = 1$, which leads to a family of infinitely many ground states $\mathbf{F_g}$ expressed as[79]

$$\mathbf{F_g} = \frac{1}{(\det\mathbf{l_d})^{1/6}}\mathbf{R_2}\mathbf{l_d}^{\frac{1}{2}}\mathbf{R_1}, \tag{26}$$

where $\mathbf{R_1}$ and $\mathbf{R_2}$ are two arbitrary rotation tensors applied in the reference and current states, respectively. Without loss of generality, we take $\mathbf{l_d}$ as the diagonal shape tensor excluding all rotations

$$\mathbf{l_d} = \begin{bmatrix} 1-Q & & \\ & 1-Q & \\ & & 1+2Q \end{bmatrix}, \tag{27}$$

with its director corresponding to $\mathbf{n_d} = \mathbf{e_3} = (0,0,1)^T$. Consequently, Eq. 26 can be rewritten as

$$\mathbf{F_g} = \mathbf{R_2 U_g R_1}, \tag{28}$$

where

$$\mathbf{U_g} = \begin{bmatrix} r^{-1/6} & & \\ & r^{-1/6} & \\ & & r^{1/3} \end{bmatrix}, \tag{29}$$

is the spontaneous deformation gradient tensor in its principal coordinates, with the chain-anisotropy ratio $r = (1+2Q)/(1-Q)$ defined before, recovering the scalar form in Eq. 13.

Each ground-state deformation in Eq. 28 can be understood by decomposing the deformation gradient $\mathbf{F_g}$ into steps illustrated in Fig. 6. Based on this kinematic picture, the nematic director $\mathbf{n_g}$ corresponding to $\mathbf{F_g}$ in the current state is expressed as

$$\mathbf{n_g} = \mathbf{R_2 n_d}. \tag{30}$$

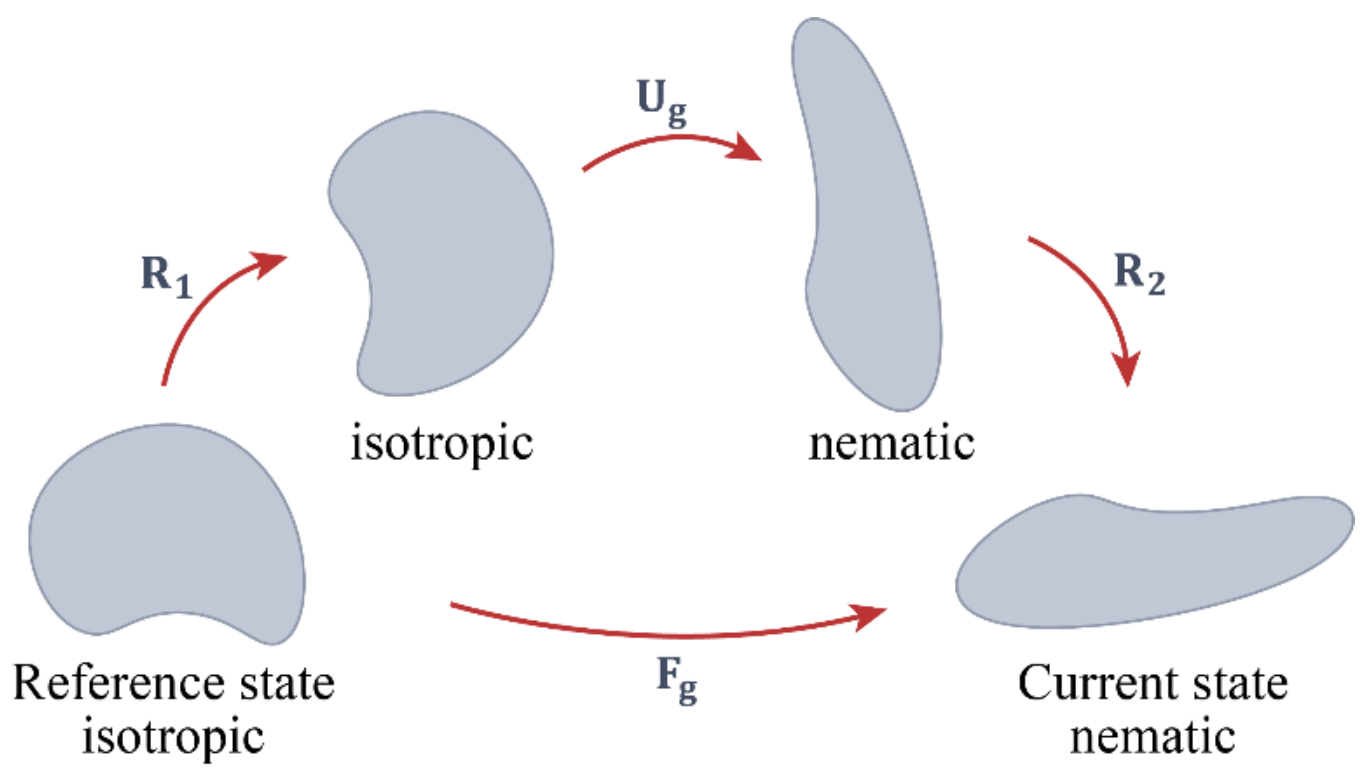

Fig. 6. Decomposition of the ground-state deformation gradient $\mathbf{F_g}$ into two arbitrary rotation tensors $\mathbf{R_1}$ and $\mathbf{R_2}$, and a spontaneous deformation gradient tensor $\mathbf{U_g}$ in its principal coordinates.

### 3.5 Stripe domains and soft elasticity

In addition to the family of infinitely many ground states $\mathbf{F_g}$ in Eq. 28, the nematic LCE can access further effective ground states by forming fine textures. One prominent example is a sheet of LCE with microscopic stripe domains under uniaxial stretch. In the following, we construct such stripe domains and the corresponding stress-stretch response.

Consider two ground states $\mathbf{F_g^+}$ and $\mathbf{F_g^-}$ in Eq. 28 expressed as

$$\mathbf{F_g^\pm} = \mathbf{R_2^\pm U_g R_1^\pm}, \tag{31}$$

where the two planar rotation tensors are defined in the $\mathbf{e_1}$-$\mathbf{e_3}$ plane as

$$\mathbf{R_1^\pm} = \begin{bmatrix} \cos\psi & 0 & \mp\sin\psi \\ 0 & 1 & 0 \\ \pm\sin\psi & 0 & \cos\psi \end{bmatrix}, \mathbf{R_2^\pm} = \begin{bmatrix} \cos\phi & 0 & \mp\sin\phi \\ 0 & 1 & 0 \\ \pm\sin\phi & 0 & \cos\phi \end{bmatrix}, \tag{32}$$

with rotation angles $\pm\psi$ and $\pm\phi$ about $-\mathbf{e_2}$. Substituting Eqs. 29 and 32 into Eq. 31, we have

$$\mathbf{F}_\mathbf{g}^{\pm} = \begin{bmatrix} r^{-1/6}\cos\psi\cos\phi - r^{1/3}\sin\psi\sin\phi & 0 & \mp\left(r^{-1/6}\sin\psi\cos\phi + r^{1/3}\cos\psi\sin\phi\right) \\ 0 & r^{-1/6} & 0 \\ \pm\left(r^{-1/6}\cos\psi\sin\phi + r^{1/3}\sin\psi\cos\phi\right) & 0 & r^{1/3}\cos\psi\cos\phi - r^{-1/6}\sin\psi\sin\phi \end{bmatrix}, \quad (33)$$

with their corresponding directors $\mathbf{n}_\mathbf{g}^{\pm} = (\mp\sin\phi, 0, \cos\phi)^T$ from Eq. 30.

For any $\phi \in [0, \pi/2]$, we can always find a corresponding $\psi$ such that $F_{\mathrm{g},31}^{+} = F_{\mathrm{g},31}^{-} = 0$, by satisfying

$$r^{-1/6}\cos\psi\sin\phi + r^{1/3}\sin\psi\cos\phi = 0. \quad (34)$$

Meanwhile, $F_{\mathrm{g},13}^{+} = -F_{\mathrm{g},13}^{-} \neq 0$ when $\phi \in (0, \pi/2)$. That is, both ground states $\mathbf{F}_\mathbf{g}^{+}$ and $\mathbf{F}_\mathbf{g}^{-}$ in this case are simple shear. The two ground states further satisfy

$$\mathbf{F}_\mathbf{g}^{+} - \mathbf{F}_\mathbf{g}^{-} = \mathbf{a} \otimes \mathbf{N}_\mathbf{int}, \quad (35)$$

with $\mathbf{a} = \left(-2\left(r^{-1/6}\sin\psi\cos\phi + r^{1/3}\cos\psi\sin\phi\right), 0, 0\right)^T$ and $\mathbf{N}_\mathbf{int} = \mathbf{e}_3 = (0,0,1)^T$.

Eq. 35 is Hadamard's compatibility condition required at the interface between two domains with compatible deformation gradients $\mathbf{F}_\mathbf{g}^{+}$ and $\mathbf{F}_\mathbf{g}^{-}$[82-84], where $\mathbf{N}_\mathbf{int} = \mathbf{e}_3$ is the normal vector of the interface. In other words, we can construct additional effective ground states by forming stripe domains of alternating layers of $\mathbf{F}_\mathbf{g}^{+}$ and $\mathbf{F}_\mathbf{g}^{-}$ (Fig. 7a). The average deformation gradient of the bulk material is then

$$\bar{\mathbf{F}} = \frac{1}{2}\left(\mathbf{F}_\mathbf{g}^{+} + \mathbf{F}_\mathbf{g}^{-}\right) = \begin{bmatrix} r^{-1/6}\cos\psi\cos\phi - r^{1/3}\sin\psi\sin\phi & & \\ & r^{-1/6} & \\ & & r^{1/3}\cos\psi\cos\phi - r^{-1/6}\sin\psi\sin\phi \end{bmatrix}. \quad (36)$$

The average free energy is

$$\bar{W} = \frac{1}{2}\left(W\left(\mathbf{F}_\mathbf{g}^{+}, \mathbf{n}_\mathbf{g}^{+}\right) + W\left(\mathbf{F}_\mathbf{g}^{-}, \mathbf{n}_\mathbf{g}^{-}\right)\right). \quad (37)$$

Because both $\mathbf{F}_\mathbf{g}^{+}$ and $\mathbf{F}_\mathbf{g}^{-}$ are ground states, $\bar{W}$ is equal to the free energy of the ground state, and lower than that of the corresponding homogeneous deformation, $W(\bar{\mathbf{F}}, \mathbf{n})$ (see Fig. 7b as a schematic illustrating the free energy landscape in 2D). Each stripe domain has non-zero shear strain $F_{\mathrm{g},13}^{\pm} \neq 0$ (except for $\phi = 0$ or $\phi = \pi/2$), but the macroscopic average deformation has zero shear strain $\bar{F}_{13} = 0$.

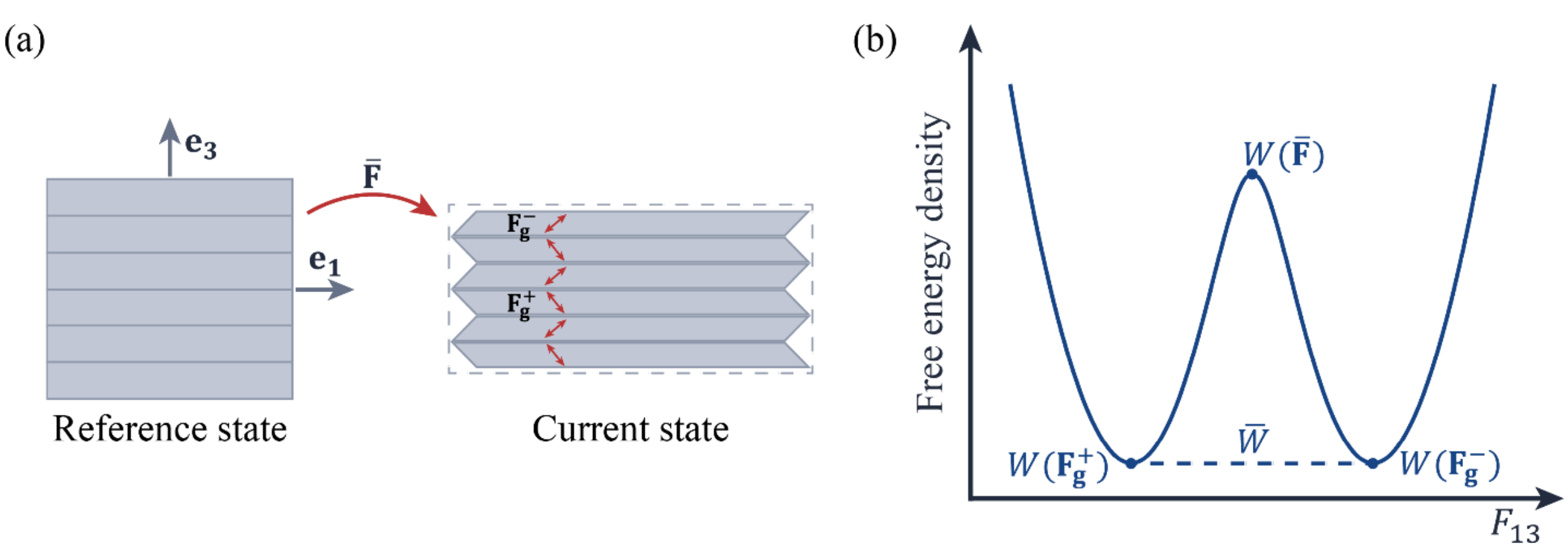

Fig. 7. Construction of additional effective ground states by forming stripe domains. (a) Stripe domains of alternating layers of $\mathbf{F}_\mathbf{g}^+$ and $\mathbf{F}_\mathbf{g}^-$ under the average deformation gradient of the bulk material, $\bar{\mathbf{F}} = \left(\mathbf{F}_\mathbf{g}^+ + \mathbf{F}_\mathbf{g}^-\right)/2$. Each stripe domain has non-zero shear strain $F_{\mathrm{g},13}^{\pm} \neq 0$ (except for $\phi = 0$ or $\phi = \pi/2$), but the macroscopic average deformation has zero shear strain $\bar{F}_{13} = 0$. (b) Schematic illustrates the free energy landscape in 2D. The free energy of homogeneous deformation $\mathbf{F}$ (solid line) shows a double-well structure with respect to $F_{13}$, where $W(\bar{\mathbf{F}})$ corresponds to the peak at $F_{13} = 0$. The homogeneous ground states $\mathbf{F}_\mathbf{g}^+$ and $\mathbf{F}_\mathbf{g}^-$ correspond to the two minima (solid dots) with nonzero shear strain $F_{\mathrm{g},13}^{\pm} \neq 0$. The average free energy $\bar{W} = \left(W\left(\mathbf{F}_\mathbf{g}^+, \mathbf{n}_\mathbf{g}^+\right) + W\left(\mathbf{F}_\mathbf{g}^-, \mathbf{n}_\mathbf{g}^-\right)\right)/2$ corresponds to the common tangent of the two minima (dashed line) and is equal to the free energy of the ground state, lower than $W(\bar{\mathbf{F}})$. To quantitatively plot such an energy landscape with respect to $F_{13}$ in 2D, we can prescribe a principal stretch, impose zero shear strain $F_{31} = 0$, and minimize the free energy over all $\mathbf{n}$ in 2D.

To further visualize the physical meanings of $\bar{\mathbf{F}}$, $\mathbf{F}_\mathbf{g}^{\pm}$, and $\mathbf{n}_\mathbf{g}^{\pm}$, let

$$\lambda = \bar{\mathrm{F}}_{11} = r^{-1/6}\cos\psi\cos\phi - r^{1/3}\sin\psi\sin\phi, \tag{38}$$

as the normal stretch along $\mathbf{e_1}$. Combining Eqs. 34, 36, and 38, we obtain

$$\bar{\mathbf{F}} = \begin{bmatrix} \lambda & & \\ & r^{-1/6} & \\ & & \dfrac{r^{1/6}}{\lambda} \end{bmatrix}, \tag{39}$$

$$\sin^2\phi = \frac{r}{r-1}\left(1 - \frac{r^{-1/3}}{\lambda^2}\right), \tag{40}$$

and

$$F_{\mathrm{g},13}^{\pm} = \mp\frac{1}{\lambda}\sqrt{(\lambda^2 - r^{-1/3})(r^{2/3} - \lambda^2)}. \tag{41}$$

From Eq. 39, we have $\det\bar{\mathbf{F}} = 1$, satisfying material incompressibility. Further, based on Eqs. 40 and 41, a zero-stress ground state always exists between $r^{-1/6} \leq \lambda \leq r^{1/3}$.

Eqs. 39-41 describe the well-known soft-elastic behavior with the formation of stripe domains in a monodomain LCE[8, 11, 85]. We still take the stress-free, undeformed isotropic phase as the reference state, and consider a square sheet of elastomer of unit length in the $\mathbf{e_1}$-$\mathbf{e_3}$ plane (Fig. 8a). In the low-temperature nematic phase, the stress-free sheet becomes monodomain with the director along $\mathbf{e_3}$. Due to thermal-induced spontaneous deformation described in Eq. 13, the sheet becomes rectangular, with stress-free stretches $\lambda_\perp = r^{-1/6}$ along $\mathbf{e_1}$, $\lambda_\parallel = r^{1/3}$ along $\mathbf{e_3}$, and $\lambda_\perp = r^{-1/6}$ in the out-of-plane direction along $\mathbf{e_2}$ (Fig. 8b).

We then stretch the monodomain sheet along $\mathbf{e_1}$, perpendicular to the initial director along $\mathbf{e_3}$. The uniaxial stretch $\lambda$ starts from $\lambda = \lambda_\perp = r^{-1/6}$. Based on Eqs. 39-41, from $\lambda = r^{-1/6}$ to $\lambda = r^{1/3}$, the sample can always stay in its stress-free ground state, with the macroscopic deformation gradient described by $\bar{\mathbf{F}}$ in Eq. 39. Consequently, between $r^{-1/6} \leq \lambda \leq r^{1/3}$, the sample is subjected to finite tensile stretch but zero tensile stress, a phenomenon called *soft elasticity*. Meanwhile, the sheet forms stripe domains of alternating layers of $\mathbf{F}_\mathbf{g}^{\pm}$ with director angles of $\pm\phi$ from $\mathbf{e_3}$ (Fig. 8c). During stretch, the director rotates from $\phi = 0$ (along $\mathbf{e_3}$) at $\lambda = r^{-1/6}$ to $\phi = \pi/2$ at $\lambda = r^{1/3}$ (along $\mathbf{e_1}$, Fig. 8d). Each stripe has nonzero shear strain $F_{g,13}^{\pm}$, but the macroscopic deformation $\bar{\mathbf{F}}$ has no shear strain, satisfying the boundary conditions of uniaxial tension. Thermodynamically, stretching within the soft-elastic regime rotates the anisotropic chain-conformation distribution—carried along by the mesogen rotation—without distorting it, so that the chain conformational entropy remains unchanged[5, 18].

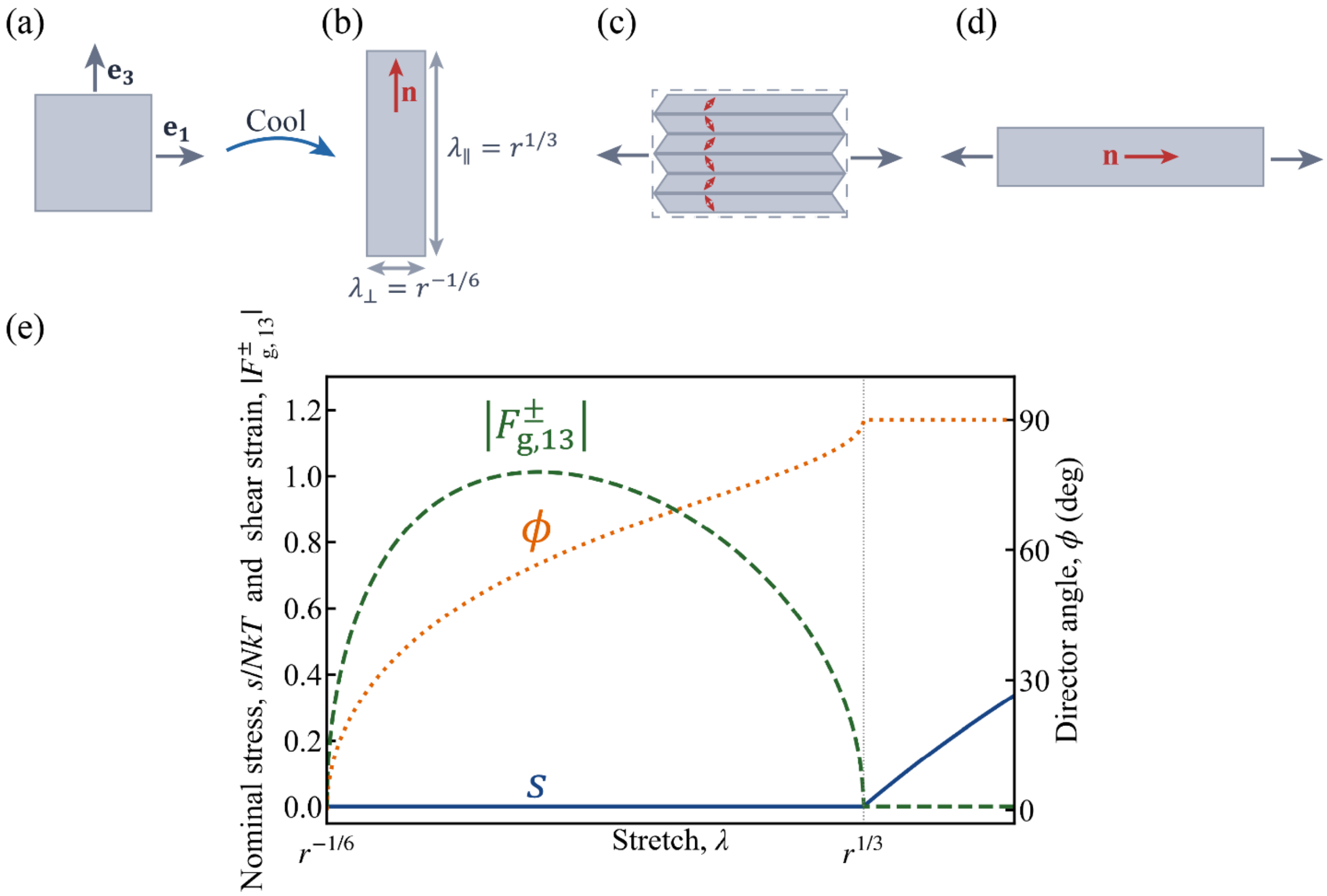

Fig. 8. Soft elasticity and the formation of stripe domains in a monodomain LCE uniaxially stretched perpendicular to its initial director. (a) The stress-free, undeformed isotropic phase is taken as the reference state. In the reference state, consider a square sheet of elastomer of unit length in the $\mathbf{e_1}$-$\mathbf{e_3}$ plane. (b) In the low-temperature nematic phase, the stress-free sheet becomes monodomain with the director along $\mathbf{e_3}$. The sheet becomes rectangular with stress-free stretches $\lambda_{\perp} = r^{-1/6}$ along $\mathbf{e_1}$ and $\lambda_{\parallel} = r^{1/3}$ along $\mathbf{e_3}$. (c) Under further stretch between $r^{-1/6} \leq \lambda \leq r^{1/3}$, the sheet forms stripe domains of alternating layers of $\mathbf{F}_\mathbf{g}^{\pm}$ with director angles of $\pm\phi$ from $\mathbf{e_3}$. (d) Afterward, the director rotates to $\phi = \pi/2$ at $\lambda = r^{1/3}$, and the sheet behaves like a common elastomer at $\lambda \geq r^{1/3}$. (e) Evolutions of the nominal stress $s$, the magnitude of shear strain $|F_{g,13}^{\pm}|$ in an individual stripe, and the director angle $\phi$, calculated analytically from Eqs. 40, 41, and 43, by assuming a constant $r = 5.5$. The vertical dashed line corresponding to $\lambda = r^{1/3}$ marks the end of soft elasticity.

When $\lambda > r^{1/3}$, the director reaches $\phi = \pi/2$, aligned with the stretching direction and does not rotate further. In this regime, no ground state is available, and the stress-stretch response is described by Eq. 11 assuming no rotation. Substituting $\lambda_3 = \lambda$ and $\lambda_1 = 1/\sqrt{\lambda}$ into Eq. 11, the free energy at $\lambda > r^{1/3}$ is

$$W(\lambda; Q) = \frac{NkT}{2}\left[\frac{\lambda^2}{1+2Q} + \frac{2\lambda^{-1}}{1-Q} + \log[(1-Q)^2(1+2Q)]\right], \qquad \lambda > r^{1/3}. \tag{42}$$

The nominal stress along the uniaxial stretch $\lambda$ is

$$s = \frac{\partial W(\lambda; Q)}{\partial \lambda} = NkT\left(\frac{\lambda}{1+2Q} - \frac{\lambda^{-2}}{1-Q}\right), \qquad \lambda > r^{1/3}. \tag{43}$$

When $Q = 0$ in the isotropic phase, the chain-anisotropy ratio $r = 1$, the soft-elastic plateau $r^{-1/6} \leq \lambda \leq r^{1/3}$ vanishes, and Eq. 43 recovers the isotropic neo-Hookean model.

Fig. 8e illustrates the evolutions of the nominal stress $s$, the shear strain $F_{g,13}^{+}$ in an individual stripe, and the director angle $\phi$, calculated analytically from Eqs. 40, 41, and 43, by assuming a constant $r = 5.5$.

To summarize, when a monodomain sample is uniaxially stretched perpendicular to its initial director, the free energy is

$$W(\lambda; Q) = \begin{cases} \dfrac{NkT}{2}\left[\dfrac{3r^{-1/3}}{1-Q} + \log[(1-Q)^2(1+2Q)]\right], & r^{-1/6} \leq \lambda \leq r^{1/3} \\ \dfrac{NkT}{2}\left[\dfrac{\lambda^2}{1+2Q} + \dfrac{2\lambda^{-1}}{1-Q} + \log[(1-Q)^2(1+2Q)]\right], & \lambda > r^{1/3} \end{cases}, \tag{44}$$

the nominal stress is

$$s = \begin{cases} 0, & r^{-1/6} \leq \lambda \leq r^{1/3} \\ NkT\left(\dfrac{\lambda}{1+2Q} - \dfrac{\lambda^{-2}}{1-Q}\right), & \lambda > r^{1/3} \end{cases}, \tag{45}$$

and the director angle $\pm\phi$ in each stripe domain is governed by

$$\sin^2\phi = \begin{cases} \dfrac{r}{r-1}\left(1 - \dfrac{r^{-1/3}}{\lambda^2}\right), & r^{-1/6} \leq \lambda \leq r^{1/3} \\ 1, & \lambda > r^{1/3} \end{cases}. \tag{46}$$

The infinite number of ground states in nematic monodomain LCEs is a direct consequence of the non-quasi-convexity of the Helmholtz free energy in Eq. 22. Mathematically, the resulting effective ground state formed by stripe domains is the quasi-convex hull of the monodomain minimizers, which will be further generalized to describe the homogenized effective behavior of polydomain LCEs (Section 4). The width of each stripe domain is determined by the competition between interfacial (Frank) energy induced by director distortion, entropic elasticity, and other sample-scale constraints, discussed in Section 5.2.

### 3.6 Ideal liquid crystal elastomers

We call an LCE that exhibits such a perfect zero-stress soft-elastic response an *ideal liquid crystal elastomer*. This idealization rests on two assumptions. First, the polymer chains are Gaussian, albeit anisotropic, so the free energy is purely entropic and excludes inelastic effects such as viscoelasticity or damage. Second, the network retains no memory of the mesogen order present at crosslinking, so the free energy only includes the current anisotropic chain conformation, in the form of $\mathbf{F}^T\mathbf{l}^{-1}\mathbf{F}$ (Eq. 22). Consequently, rotating the director in the current state costs no increase of free energy, giving a continuous family of ground states and perfect soft elasticity. Real LCEs depart from this idealization: depending on their synthesis and processing, they typically behave semi-soft or non-soft, which requires additional nonideality terms, as discussed in Section 5.3.

**3.7 Monodomain phase as a reference state**

We may also adopt the stress-free nematic monodomain state as the reference state and define the deformation gradient $\mathbf{F}$ in the current state with respect to it. To do so, we take $Q_0$ and $\mathbf{n}_0$ as the order parameter and director in the reference state (stress-free nematic monodomain), and take $Q$ and $\mathbf{n}$ as those in the current state. Following the same freely jointed chain model but using the new reference state, the Helmholtz free energy takes the form[4, 18]

$$W(\mathbf{F}, \mathbf{n}; Q, Q_0, \mathbf{n}_0) = \frac{NkT}{2}\left[\mathrm{Tr}(\mathbf{l}_0\mathbf{F}^T\mathbf{l}^{-1}\mathbf{F}) + \log\left(\frac{\det\mathbf{l}}{\det\mathbf{l}_0}\right)\right], \tag{47}$$

where

$$\mathbf{l}_0 = (1 - Q_0)\boldsymbol{\delta} + 3Q_0\mathbf{n}_0 \otimes \mathbf{n}_0, \tag{48}$$

is the shape tensor in the reference state. In this case, the soft-elastic regime discussed in Section 3.5 becomes $1 \le \lambda \le r^{1/2}$, where the previous stress-free $\lambda = r^{-1/6}$ is taken as the new reference state $\lambda = 1$.

Setting $Q_0 = 0$ and $\mathbf{l}_0 = \boldsymbol{\delta}$ in the isotropic phase, Eq. 47 recovers the original Eq. 22. In the remainder of this paper, we will keep the stress-free, undeformed isotropic phase as the reference state and base our subsequent discussions on Eq. 22.

**3.8 Free energy using normalized shape tensor**

The shape tensor $\mathbf{l}$ in Eq. 22 is not isochoric, i.e., $\det(\mathbf{l}) \neq 1$. For mathematical convenience, we may rewrite Eq. 22 using a normalized shape tensor $\tilde{\mathbf{l}}$ with $\det(\tilde{\mathbf{l}}) = 1$. To do so, we express the isochoric tensor $\tilde{\mathbf{l}}$ as

$$\tilde{\mathbf{l}} = \frac{1}{(\det\mathbf{l})^{1/3}}\mathbf{l} = \frac{r^{-1/3}}{(1 - Q)}\mathbf{l} = r^{-1/3}[\boldsymbol{\delta} + (r - 1)\mathbf{n} \otimes \mathbf{n}]. \tag{49}$$

Substituting Eq. 49 into Eq. 22, we obtain

$$W(\mathbf{F}, \mathbf{n}; Q) = \frac{NkTr^{-1/3}}{2(1 - Q)}\left[\mathrm{Tr}\left(\mathbf{F}^T\tilde{\mathbf{l}}^{-1}\mathbf{F}\right) + (1 - Q)r^{1/3}\log[(1 - Q)^2(1 + 2Q)]\right], \tag{50}$$

where the effective right Cauchy-Green tensor $\mathbf{F}^T\tilde{\mathbf{l}}^{-1}\mathbf{F}$ is isochoric.

We may further replace the deformation-independent term $(1-Q)r^{1/3}\log[(1-Q)^2(1+2Q)]$ with a constant $-3$, such that $W(\mathbf{F},\mathbf{n};Q)=0$ in the ground state, without affecting the relationship between stress and deformation gradient. The modified free energy is

$$W(\mathbf{F},\mathbf{n};Q)=\frac{NkTr^{-1/3}}{2(1-Q)}\left[\mathrm{Tr}\left(\mathbf{F}^T\tilde{\mathbf{I}}^{-1}\mathbf{F}\right)-3\right]. \tag{51}$$

To be consistent with most literature, Eq. 51 will be taken as the basis of the quasi-convex free energy for polydomain LCEs discussed next (Section 4).

## 4. Polydomain LCEs

We have shown that because of the infinitely many ground states, a stress-free nematic LCE can comprise many geometrically compatible, coexisting monodomains, typically at the microscale, each with its own director $\mathbf{n}$ but the same order parameter $Q(T)$ governed primarily by temperature. Such a sample is called a *polydomain* LCE, analogous to a polycrystal in metals. If the distribution of $\mathbf{n}$ among domains is completely random, the polydomain LCE has isotropic properties at the macroscale and shows no spontaneous deformation upon temperature change, in contrast to the monodomain discussed in Section 3. The microscale domains with different directors scatter light, making the polydomain LCE appear opaque in the nematic state, in contrast to the transparent monodomain sample and the high-temperature isotropic phase (Fig. 9). This optical difference serves as a useful way to differentiate polydomain, monodomain, and isotropic LCEs. Upon uniaxial stretching, a polydomain LCE exhibits soft elasticity: mesogens in each compatible ground-state monodomain progressively rotate toward the stretching direction, until all domains coalesce into a single director aligned with the stretch and the whole sample becomes monodomain[86, 87].

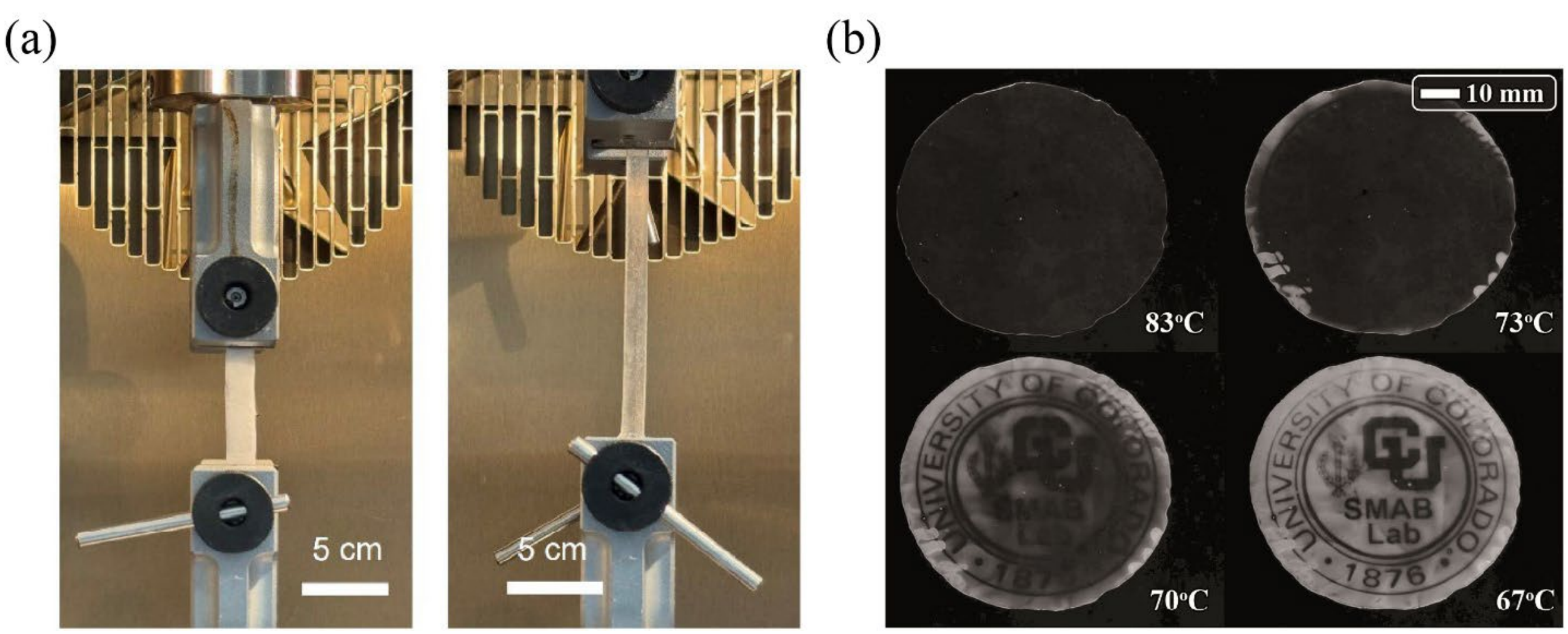

Fig. 9. Optical properties of polydomain, monodomain, and isotropic LCEs. (a) A nematic ($T<T_{\mathrm{ni}}$) unstretched polydomain LCE is opaque. Under uniaxial stretch, all mesogens are eventually aligned toward a single director, forming a monodomain LCE that becomes transparent, still in the nematic phase ($T<$

$T_{\mathrm{ni}}$). Reproduced with permission.[27] (b) A polydomain LCE is in the isotropic phase and is transparent above $T_{\mathrm{ni}}$ (top, 83 °C and 73 °C). The same LCE is in the nematic phase and is opaque below $T_{\mathrm{ni}}$ (bottom, 70 °C and 67 °C). Reproduced with permission.[19]

Mechanics of polydomain LCEs can be modeled numerically using the same monodomain free energy (Eq. 22 or Eq. 51), considering heterogeneous domain distribution[25, 88]. However, because this free energy function is non-quasi-convex, its direct minimization is ill-posed. Additional terms must therefore be introduced as a regularizer to restore well-posedness. A common choice is the Frank energy (Section 5.2), which acts as an effective interfacial energy quadratic in $\nabla\mathbf{n}$ and penalizes director distortion across domains. While this approach captures the multiscale material behavior by resolving the detailed domain-level mesogen orientation, it can be computationally costly, especially for multi-dimensional loads, complex loading paths, or large samples[88]. Alternatively, as first proposed by Conti, DeSimone, Dolzmann et al.[11-14], one may treat the polydomain sample as a homogenized continuum without resolving the subdomain processes, and describe the macroscopic material response using an effective energy as the quasi-convex relaxation of the monodomain free energy, minimized over all admissible fine-scale director and deformation fields. In the following, we present this quasi-convex free energy of polydomain LCEs, together with its recent extensions (Sections 4.1 and 4.2), within the same theoretical framework and material parameters introduced so far. Using this theory, we show the uniaxial, equal-biaxial, and non-equal-biaxial responses of a polydomain LCE, together with their physical implications (Section 4.3).

### 4.1 Continuum framework

Consider a stress-free, undeformed polydomain LCE with a completely random distribution of domain-level directors (Fig. 10), which we set as the reference state. Because of the random distribution of directors, the sample shows no spontaneous deformation upon temperature change, so that we may assign an arbitrary temperature to the reference state without affecting the subsequent kinematic relations. Following the homogenization approach discussed above, each material particle $\mathbf{X}$ now represents a collection of many monodomains. Separation of length scales requires the size of the material particle $\mathbf{X}$ to be much larger than the individual domain size, ensuring a full statistical representation of the randomly distributed directors. In the current state at temperature $T$, the LCE has an order parameter $Q(T)$, deforms to a shape defined by a field of current particle positions $\mathbf{x}(\mathbf{X})$, while its director field is averaged out at the material-particle level. The deformation gradient is defined in the same way $\mathbf{F} = \partial\mathbf{x}/\partial\mathbf{X}$, with material incompressibility $\det\mathbf{F} = 1$.

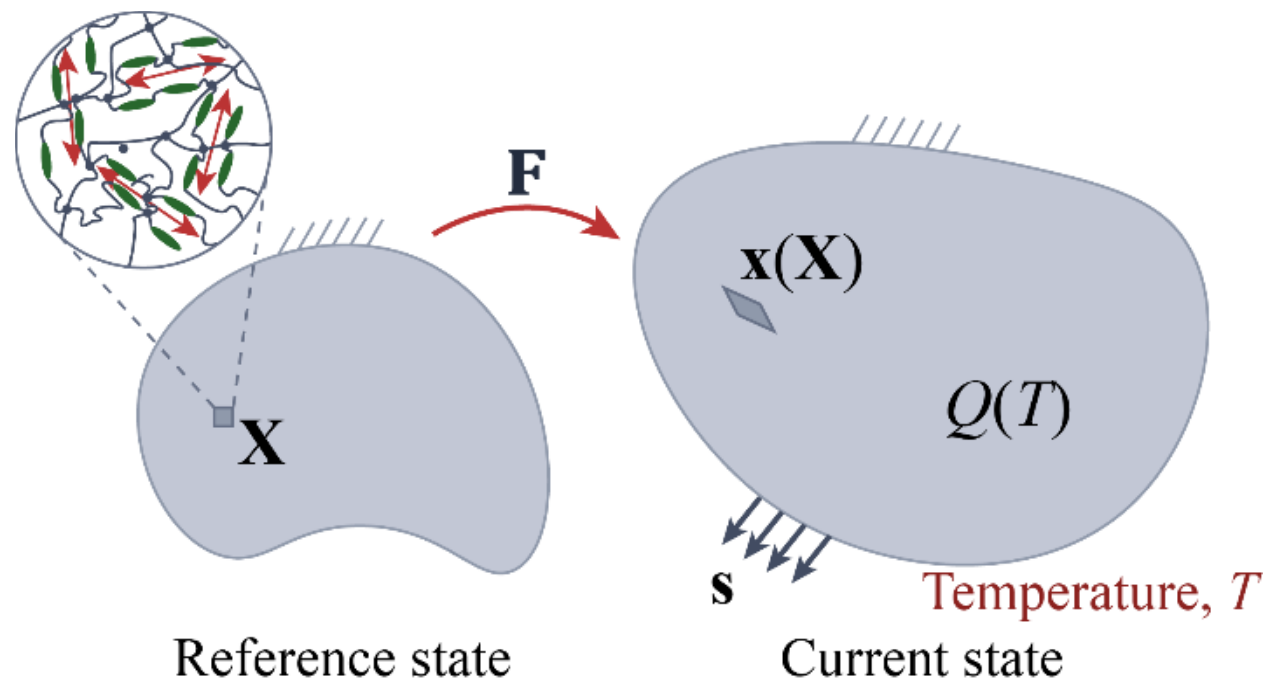

Fig. 10. Continuum framework of a polydomain LCE. (a) In the reference state, a stress-free, undeformed polydomain LCE has a completely random distribution of domain-level directors. Each material particle **X** represents a collection of many monodomains with a full statistical representation of the randomly distributed directors. The reference state can be assigned with an arbitrary temperature without affecting the kinematics of the theory. (b) In the current state at temperature $T$, under a field of nominal stress **s**, the LCE has an order parameter $Q(T)$ and deforms to a shape defined by a field of current particle positions $\mathbf{x}(\mathbf{X})$, while its director field is averaged out at the material-particle level.

### 4.2 Quasi-convex free energy and mechanical phases

The effective mechanical behavior of such a homogenized material particle in polydomain LCEs is described by the quasi-convex free energy, first developed by Conti, DeSimone, Dolzmann et al.[11-14] as an extension of the monodomain neo-classical model and later generalized by Lee, Wihardja, and Bhattacharya[89]. Here we present a version of this model whose underlying statistical description is consistent with that of the monodomain neo-classical model (Eqs. 6-9)[90].

To be consistent with the literature, we start from the monodomain neo-classical model in the form of Eq. 51. Its quasi-convex envelope over all admissible fine-scale director and deformation fields is expressed as

$$W_{\mathrm{qc}}(\lambda_1, \lambda_2, \lambda_3; Q) = \begin{cases} 0, & \text{Phase SE: } \lambda_1 \geq r^{-1/6} \\ \dfrac{NkTr^{-1/3}}{2(1-Q)}(\tilde{I}_1 - 3), & \text{else} \end{cases}, \tag{52}$$

where $\tilde{I}_1$ is the relaxed first invariant of the isochoric effective right Cauchy-Green tensor $\mathbf{F}^T\tilde{\mathbf{I}}^{-1}\mathbf{F}$ in Eq. 51, expressed as

$$\tilde{I}_1(\lambda_1, \lambda_2, \lambda_3; Q) = \begin{cases} r^{1/3}(\lambda_3^2 r^{-1} + \lambda_2^2 + \lambda_1^2), & \text{Phase M: } r^{-1/2}\lambda_3^2\lambda_1 > 1 \\ r^{1/3}(\lambda_1^2 + 2\lambda_1^{-1}r^{-1/2}), & \text{Phase P: else} \end{cases}, \tag{53}$$

and the three principal stretches satisfy $\lambda_3 \geq \lambda_2 \geq \lambda_1$, subject to incompressibility $\lambda_1\lambda_2\lambda_3 = 1$.

The quasi-convex elastic energy in Eqs. 52 and 53 defines three "phases" that describe the characteristics of their associated macroscopic mechanical responses under arbitrary biaxial stretches. These correspond to the sets $K^{qc}$, $I$, and $S$ in Biggins et al. (2012)[16]—zero-energy, energy depending only on the smallest principal stretch, and energy depending on all three—which we relabel as Phases *SE*, *P*, and $M$[90]. A phase diagram is commonly constructed based on the areal stretch $1/\lambda_1 = \lambda_2\lambda_3$ and the largest principal stretch $\lambda_3$ (Fig. 11). The phase diagram is bounded by $1/\lambda_1 \geq \sqrt{\lambda_3}$ and $1/\lambda_1 \leq \lambda_3^2$, corresponding to the uniaxial and equal-biaxial loading paths.

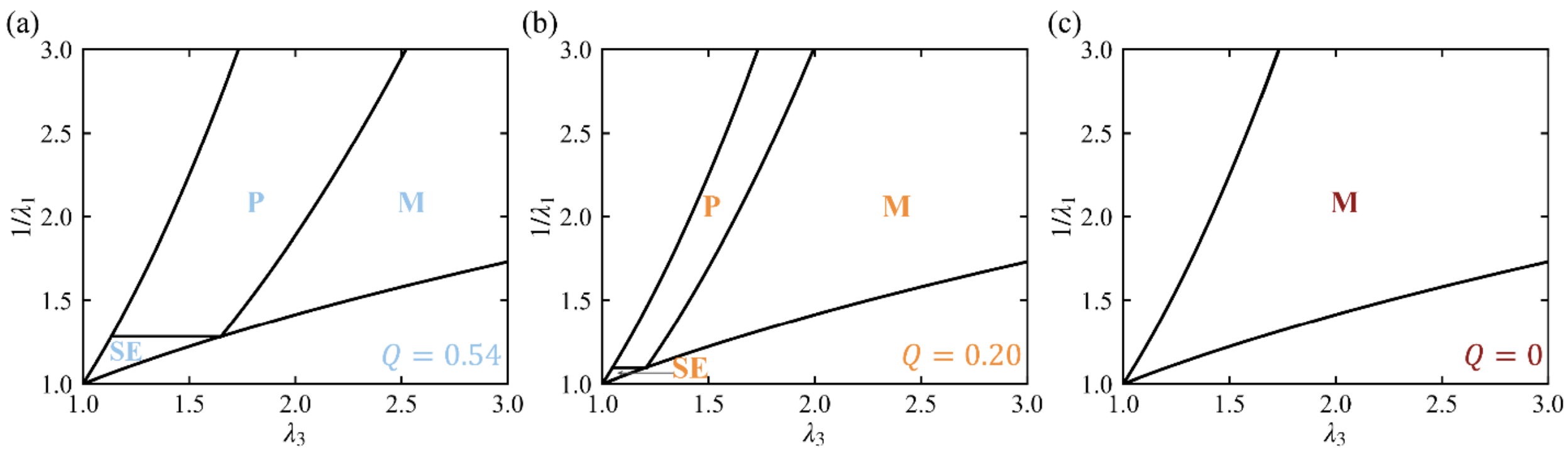

Fig. 11. Mechanical phase diagrams of a polydomain LCE at different temperatures expressed in terms of the order parameter $Q$: (a) $Q = 0.54$, (b) $Q = 0.20$, and (c) $Q = 0$. These correspond to (a) around room temperature, (b) an intermediate temperature below $T_{\mathrm{ni}}$, and (c) above $T_{\mathrm{ni}}$. The phase diagram is based on the areal stretch $1/\lambda_1 = \lambda_2\lambda_3$ and the largest principal stretch $\lambda_3$. The phase diagram is bounded by $1/\lambda_1 \geq \sqrt{\lambda_3}$ and $1/\lambda_1 \leq \lambda_3^2$, corresponding to the uniaxial and equal-biaxial loading paths. The three phases are represented by *soft elastic* (SE), *planar* (P), and *monodomain* (M).

In Phase SE (*soft elastic*), the polydomain LCE undergoes finite stretches at zero stress, accommodated by collective, domain-level mesogen rotation, a direct consequence of the infinitely many ground states of the monodomain neo-classical energy. Similar to the uniaxial soft elasticity discussed in Section 3.5, this generalized soft-elastic deformation has been reported in experiments, accompanied by the formation and reorganization of finely striped domains of mesogens[91]. In Phase P (*planar*), the LCE behaves liquid-like in-plane but solid-like out-of-plane, and the elastic energy depends only on the areal stretch $1/\lambda_1$. Correspondingly, the LCE develops more complex microstructures, including circular double laminates and multiscale, hierarchical lamination, although the mechanisms that select the specific types and scales of these microstructures remain unclear[91]. In Phase M (*monodomain*), all mesogens are aligned along a single director, such as that of an applied uniaxial stretch. The LCE becomes monodomain and behaves like a conventional rubber upon further stretching. The boundary between Phase SE and Phase P is described by $1/\lambda_1 = r^{1/6}$ and $r^{1/12} \leq \lambda_3 \leq r^{1/3}$. The boundary between Phase P and Phase M is described by $1/\lambda_1 =$ $r^{-1/2}\lambda_3^2$. Since $r = (1+2Q)/(1-Q)$ depends on $T$, the boundaries shift with the temperature change (Figs. 11a-c). For example, upon heating, Phase SE and Phase P gradually shrink and eventually disappear at high temperatures when $Q = 0$.

Lee, Wihardja, and Bhattacharya further generalized this quasi-convex formulation to various forms of free energy functions based on $\tilde{I}_1$[89]. Their generalization follows Agostiniani and DeSimone[92] in replacing the neo-Hookean form with a generalized Mooney-Rivlin energy to capture stiffening at large stretches. In the same spirit, we adopt here a Gent-like model that considers finite extensibility of polymer chains[90, 93], expressed as

$$W_{\mathrm{G}}(\lambda_1,\lambda_2,\lambda_3;Q) = \begin{cases} 0, & \text{Phase SE: } \lambda_1 \geq r^{-1/6} \\ -\dfrac{NkTr^{-1/3}J_{\mathrm{lim}}}{2(1-Q)}\log\left(1-\dfrac{\tilde{I}_1-3}{J_{\mathrm{lim}}}\right), & \text{else} \end{cases}, \tag{54}$$

where $J_{\mathrm{lim}}$ is a fitting parameter that accounts for finite chain extensibility. When $J_{\mathrm{lim}} \to \infty$, Eq. 54 reduces to the quasi-convex energy in Eq. 52. A finite $J_{\mathrm{lim}}$ introduces a terminal stiffening effect that limits the maximum stretch allowed under loading.

### 4.3 Uniaxial, equal-biaxial, and non-equal-biaxial responses

To illustrate the mechanical behavior of polydomain LCEs, we next calculate the uniaxial, equal-biaxial, and non-equal-biaxial stress-stretch responses using the Gent-like model in Eq. 54.

Considering $\lambda_3 \geq \lambda_2 \geq \lambda_1$ with $\lambda_1\lambda_2\lambda_3 = 1$, we describe the general in-plane stretch state using $\lambda_2$ and $\lambda_3$. Substituting $\lambda_1 = 1/(\lambda_2\lambda_3)$ into Eq. 54, we rewrite the incompressible Gent-like model as

$$W_{\mathrm{G}}(\lambda_2,\lambda_3;Q) = \begin{cases} 0, & \text{Phase SE: } \lambda_2\lambda_3 \leq r^{1/6} \\ -\dfrac{NkTr^{-1/3}J_{\mathrm{lim}}}{2(1-Q)}\log\left(1-\dfrac{(\tilde{I}_1-3)}{J_{\mathrm{lim}}}\right), & \text{else} \end{cases}, \tag{55}$$

where

$$\tilde{I}_1(\lambda_2,\lambda_3;Q) = \begin{cases} r^{1/3}(\lambda_2^2+\lambda_2^{-2}\lambda_3^{-2}+\lambda_3^2 r^{-1}), & \text{Phase M: } \lambda_2^{-1}\lambda_3 > r^{1/2} \\ r^{1/3}(\lambda_2^{-2}\lambda_3^{-2}+2\lambda_2\lambda_3 r^{-1/2}), & \text{Phase P: else} \end{cases}. \tag{56}$$

The principal true (Cauchy) stresses follow the relationship

$$\begin{cases} \sigma_2-\sigma_1 = \lambda_2\dfrac{\partial W_{\mathrm{G}}(\lambda_2,\lambda_3;Q)}{\partial\lambda_2} \\ \sigma_3-\sigma_1 = \lambda_3\dfrac{\partial W_{\mathrm{G}}(\lambda_2,\lambda_3;Q)}{\partial\lambda_3}, \end{cases} \tag{57}$$

and the nominal stresses follow

$$\begin{cases} s_2 = \sigma_2/\lambda_2 \\ s_3 = \sigma_3/\lambda_3 \end{cases}. \tag{58}$$

In uniaxial stretch, we take $\lambda_3 = \lambda$ as the largest principal stretch and $\lambda_1 = \lambda_2 = \lambda^{-1/2}$, with the corresponding stresses $\sigma_3 = \sigma$ and $\sigma_1 = \sigma_2 = 0$. In equal-biaxial stretch, we take $\lambda_2 = \lambda_3 = \lambda$ as the largest principal stretches and $\lambda_1 = \lambda^{-2}$, with the corresponding stresses $\sigma_2 = \sigma_3 = \sigma$. Substituting these into Eqs. 55-57, we obtain the explicit expressions of true stresses in each case:

$$\text{Uniaxial: } \sigma = \begin{cases} 0, & \text{Phase SE: } 1 \le \lambda \le r^{1/3} \\ \dfrac{NkT}{1-Q} \dfrac{r^{-1}\lambda^2 - \lambda^{-1}}{1 - \dfrac{r^{1/3}(r^{-1}\lambda^2 + 2\lambda^{-1}) - 3}{J_{\lim}}}, & \text{Phase M: } \lambda > r^{1/3} \end{cases}, \tag{59}$$

and

$$\text{Equal-biaxial: } \sigma = \begin{cases} 0, & \text{Phase SE: } 1 \le \lambda \le r^{1/12} \\ \dfrac{NkT}{1-Q} \dfrac{r^{-1/2}\lambda^2 - \lambda^{-4}}{1 - \dfrac{r^{1/3}(\lambda^{-4} + 2r^{-1/2}\lambda^2) - 3}{J_{\lim}}}, & \text{Phase P: } \lambda > r^{1/12} \end{cases}. \tag{60}$$

Fig. 12 compares the stress-stretch curves of the uniaxial and equal-biaxial states, with the soft-elastic regime spanning from $\lambda = 1$ to $\lambda = r^{1/3}$ and $r^{1/12}$, respectively. Note that the upper bound of the soft-elastic regime, $\lambda = r^{1/3}$, in the uniaxial case coincides with the monodomain spontaneous stretch $\lambda_{\parallel}$ in Eq. 13, indicating that a fully aligned stress-free polydomain LCE reproduces the monodomain counterpart.

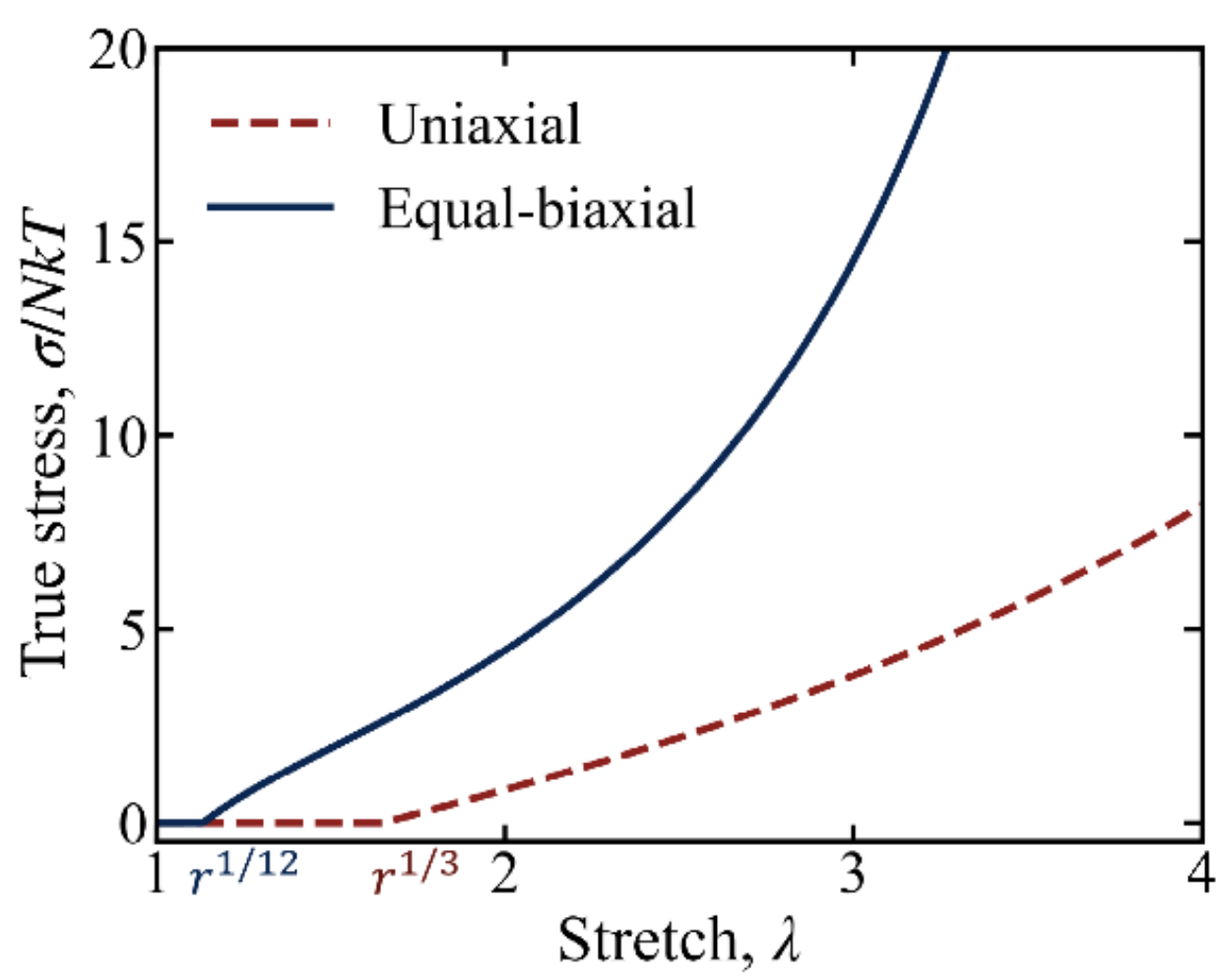

Fig. 12. Theoretically calculated uniaxial and equal-biaxial stress-stretch responses of a polydomain LCE at $Q = 0.54$, with $J_{\lim} = 30$. The upper bound of the soft-elastic regime is labeled in each case, $\lambda = r^{1/3}$ and $r^{1/12}$, respectively.

We further show a general non-equal-biaxial stretch state with $\lambda_2 \neq \lambda_3$. Considering only in-plane loads with zero out-of-plane stress $\sigma_1 = 0$, we obtain the true stresses in Phase SE and Phase P as (the loading path may further enter Phase M depending on the values of $\lambda_2$ and $\lambda_3$, which is not discussed here)

Non-equal-biaxial:

$$\sigma_2 = \sigma_3 = \begin{cases} 0, & \text{Phase SE: } \lambda_2\lambda_3 \le r^{1/6} \\ \dfrac{NkT}{1-Q}\dfrac{r^{-1/2}\lambda_1^{-1} - \lambda_1^2}{1 - \dfrac{r^{1/3}(\lambda_1^2 + 2r^{-1/2}\lambda_1^{-1}) - 3}{J_{\text{lim}}}}, & \text{Phase P: } \lambda_2\lambda_3 > r^{1/6} \text{ and } \lambda_2^{-1}\lambda_3 \le r^{1/2} \end{cases}, \quad (61)$$

where $\lambda_1^{-1} = \lambda_2\lambda_3$ is the in-plane areal stretch.

Eq. 61 shows the in-plane liquid-like behavior of a polydomain LCE observed in experiments[25]: within Phase P, the material can undergo arbitrary unequal in-plane stretches $\lambda_2 \neq \lambda_3$, but the corresponding in-plane true stresses always remain equal, $\sigma_2 = \sigma_3$. In other words, the LCE can be arbitrarily sheared in plane without inducing any shear stress, behaving like an inviscid liquid. This in-plane liquid-like behavior is a direct consequence of the quasi-convex free energy in Phase P (through $\tilde{I}_1$ in Eq. 53 or Eq. 56): it only depends on the areal stretch $\lambda_1^{-1} = \lambda_2\lambda_3$ instead of the individual stretches or the loading path. The LCE accommodates an imposed in-plane shear at no shear stress, with the directors rearranging within the plane to relieve it. Fig. 13 compares the non-equal-biaxial stress-stretch responses of a polydomain LCE to a common isotropic rubber, where the latter shows unequal in-plane stresses $\sigma_2 \neq \sigma_3$ under the prescribed unequal in-plane stretches $\lambda_2 \neq \lambda_3$.

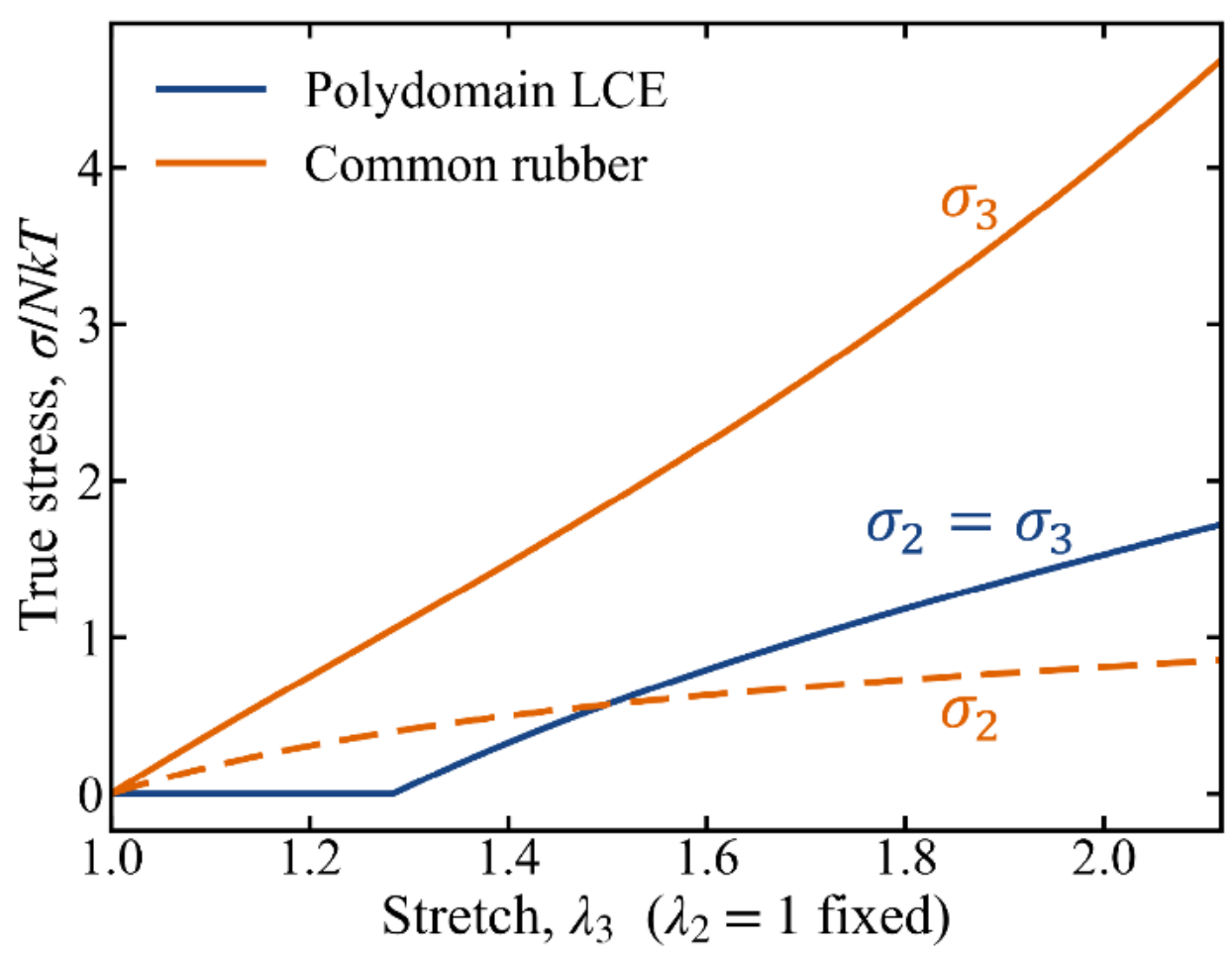

Fig. 13. Theoretically calculated non-equal-biaxial stress-stretch responses of a polydomain LCE and a common isotropic rubber. The loading protocol fixes $\lambda_2 = 1$ and increases $\lambda_3$, with the areal stretch $\lambda_1^{-1} = \lambda_2\lambda_3 = \lambda_3$. For comparison, the two materials are modeled using the same model in Eq. 55, with $Q = 0.54$ for the polydomain LCE, $Q = 0$ for the common rubber, and $J_{\text{lim}} = 30$ in both cases.

### 5. Beyond the ideal elastic theory

The ideal elastic theories in Sections 3 and 4 capture the backbone of LCE mechanics. Quantitative modeling of real materials calls for further extensions. Here we briefly discuss four such extensions, building on the same ideal framework discussed above. These extensions include the temperature- and light-dependence of LC mesogens and the resulting thermomechanical and photomechanical coupling (Section 5.1), the Frank elasticity that governs the detailed heterogeneous director field and domain microstructure (Section 5.2), the LCE network nonideality that controls genesis-dependent semi-softness and links processing to properties (Section 5.3), and the viscoelasticity that governs the rate-dependent response (Section 5.4).

**5.1 Thermomechanical and photomechanical coupling**

So far, we have treated the order parameter $Q$ as a control parameter that depends only on temperature in the free energy. Theoretically, $Q$ can instead be obtained from an additional free energy describing the mesogen interaction itself. Here we summarize two widely used descriptions of this free energy, the phenomenological Landau-de Gennes expansion and the molecular Maier-Saupe theory. We then extend the free energy to photoactive LCEs where the elastomer responds to light through photochemical reaction.

In LC melts or solutions, the thermo-responsive nematic-isotropic phase transition is weakly first-order and can be described by the phenomenological Landau-de Gennes theory[1, 18]. The theory expands the free energy in powers of the scalar order parameter $Q$ near the nematic-isotropic transition. The polynomial can be arranged in various forms to describe different material systems, with one such example written as

$$W_{\mathrm{lc}}(Q) = \frac{1}{2}A_0(T - T^*)Q^2 - \frac{1}{3}BQ^3 + \frac{1}{4}CQ^4, \tag{62}$$

where $A_0$, $B$, and $C$ are positive constants and $T^*$ is a reference temperature.

The free energy shows a double-well structure over a range of temperatures (Fig. 14a), manifesting the first-order nature of the transition between the nematic ($Q > 0$) and isotropic ($Q = 0$) phases. The temperature enters only through the quadratic term, whose coefficient $A_0(T - T^*)$ changes sign at $T^*$. Physically, $T^*$ is the temperature at which the isotropic minimum at $Q = 0$ loses its stability, marking the lower limit of supercooling of the isotropic phase. The cubic term is included to break the symmetry between $+Q$ and $-Q$, since they represent distinct physical states of rods aligned along the director and rods confined to the plane perpendicular to the director, respectively. The cubic term also makes the nematic-isotropic transition weakly first-order rather than continuous, such that at the transition temperature $T_{\mathrm{ni}}$, $Q$ jumps discontinuously from zero to a finite value $Q_{\mathrm{ni}}$, governed by[18]

$$T_{\mathrm{ni}} = T^* + \frac{2B^2}{9A_0C}, Q_{\mathrm{ni}} = \frac{2B}{3C}. \tag{63}$$

This ordered energy minimum at $Q > 0$ persists up to a superheating limit $T_{\mathrm{u}} = T^* + B^2/(4A_0C)$. Between $T^*$ and $T_{\mathrm{u}}$, the free energy is double-well with both the nematic ($Q > 0$) and isotropic ($Q = 0$) phases locally stable, which describes the observed thermal hysteresis in experiments.

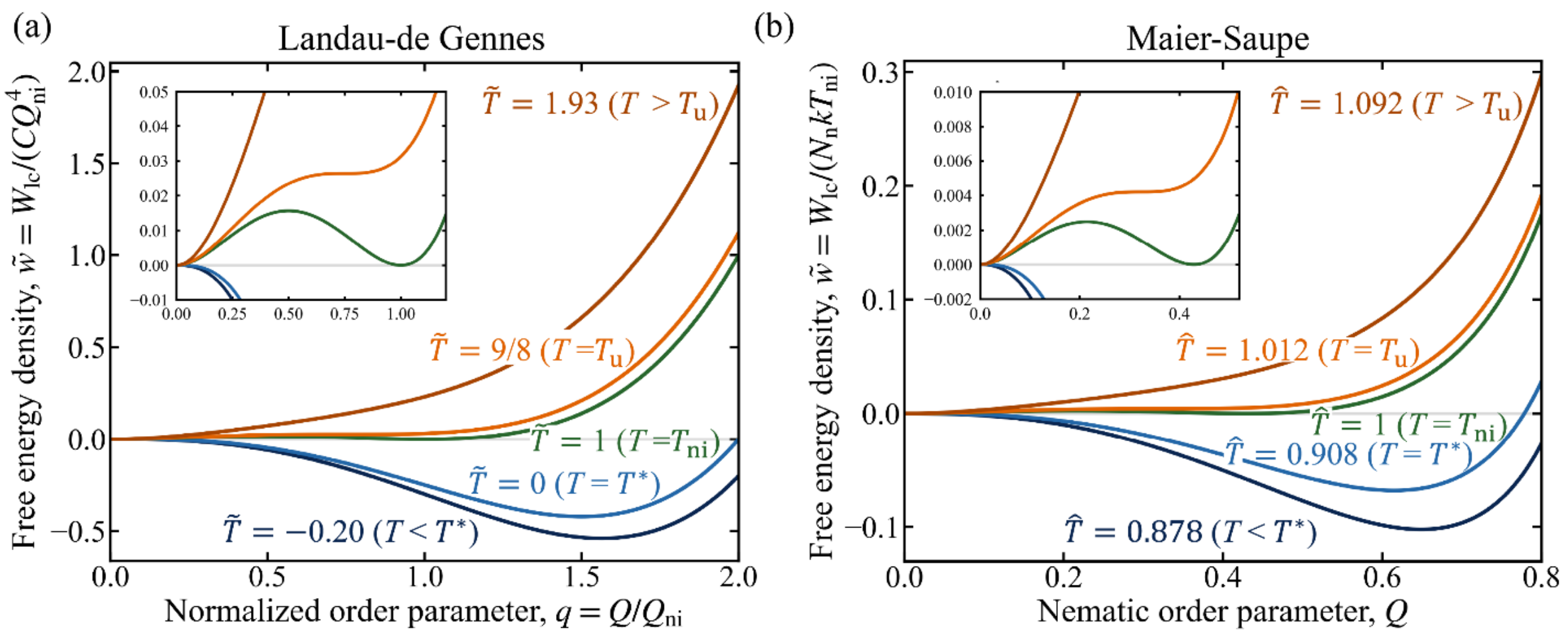

Fig. 14. Free energy landscapes describing the mesogen interaction, shown in dimensionless forms. (a) Landau-de Gennes free energy. Using the normalized order parameter $q = Q/Q_{ni}$, temperature $\tilde{T} = (T - T^*)/(T_{ni} - T^*)$, and free energy function $\tilde{w} = W_{lc}(Q)/(CQ_{ni}^4)$, Eq. 62 reduces to a dimensionless form $\tilde{w}(q) = (\tilde{T}/4)q^2 - (1/2)q^3 + (1/4)q^4$. (b) Maier-Saupe free energy. Defining the normalized temperature $\hat{T} = T/T_{ni}$ and free energy function $\tilde{w} = W_{lc}(Q)/N_n kT_{ni}$, the dimensionless interaction parameter $\hat{J} = J/kT_{ni}$ is obtained by setting $\hat{T} = 1$ (at which the free energy has two minima with equal depth at $Q = 0$ and $Q = Q_{ni}$). Solving this condition numerically gives $\hat{J} = 4.54$.

Values of $A_0$, $B$, and $C$ are typically fitted by experimental measurements, e.g., based on the following three independent observations[18]. First, a representative value of $Q_{ni} \approx 0.4$ at the transition point $T = T_{ni}$ gives $B \approx 0.6C$. Second, $T_{ni} - T^* \approx 1$ K in common thermotropic LCs sets $(2B^2)/(9A_0C) \approx 1$ K. Finally, the latent heat per unit volume, $\Delta H \approx (1/2)T_{ni}A_0Q_{ni}^2$, obtained from calorimetry, sets the overall order of magnitude. With these, commonly used $A_0$ is of order $10^5$ J/(m$^3 \cdot$ K), and $B$ and $C$ of order $10^6$-$10^7$ J/m$^3$.

Another common choice of $W_{lc}(Q)$ is the molecular description by the Maier-Saupe theory[76, 77]. This theory assumes dipolar-like long-range molecular interaction between individual mesogens, leading to a total potential energy of the system as $-(1/2)\sum_{i \neq j} J_{ij} Q_i Q_j$, where $J_{ij}$ denotes the interaction potential between the $i_{th}$ and $j_{th}$ mesogens and $Q_i$ is the corresponding second Legendre polynomial of the cosine of the angle between the $i_{th}$ mesogen and the director. The Helmholtz free energy is expressed as (see a full derivation in ref.[79])

$$W_{lc}(Q) = N_n kT \left[ g^{-1}(Q)Q - \log Z(Q) - \frac{J}{2kT} Q^2 \right], \tag{64}$$

$$g(x) = -\frac{1}{2} - \frac{1}{2x} + \frac{1}{2x} \sqrt{\frac{3x}{2}} \frac{\exp(3x/2)}{\int_0^{\sqrt{3x/2}} \exp(y^2) \mathrm{d}y}, \tag{65}$$

$$Z(Q) = \frac{\exp[g^{-1}(Q)]}{1 + g^{-1}(Q)(1 + 2Q)}, \tag{66}$$

where $N_{\mathrm{n}}$ is the number of mesogens per unit volume, $g^{-1}(Q)$ the trial field, $Z(Q)$ the single-mesogen partition function, and $J$ the average interaction strength. Minimizing Eq. 64 over $Q$ at each temperature gives the equilibrium order parameter $Q(T)$. The Maier-Saupe free energy in Eq. 64 exhibits a similar double-well structure (Fig. 14b), reproducing the weakly first-order transition. Expanding the free energy function to fourth order recovers the Landau-de Gennes form of Eq. 62, with $T^*$ and $T_{\mathrm{ni}}$ set by the interaction strength $J$. For example, at room temperature $T = 300\ \mathrm{K}$, setting $J = 5kT$ leads to $T_{\mathrm{ni}} \approx 1.1T = 330\ \mathrm{K}$ and $Q_{\mathrm{ni}} \approx 0.4$, in qualitative agreement with measurements of modern LCEs.

The Maier-Saupe theory can be further extended to describe the photomechanical behavior of photoactive LCEs. These LCEs incorporate photochromic molecules, most commonly azobenzene, in their network through mixing or linkage to polymer chains. First reported by Finkelmann et al. in 2001[17], illumination at a specific wavelength drives a photochemical reaction in which the azobenzene isomerizes from a rod-like *trans* state to a bent *cis* state. This photoisomerization dilutes the nematic population, reduces the nematic order parameter $Q$, and induces a macroscopic deformation of the LCE network.

To describe this photochemically induced deformation, Corbett and Warner extended the Maier-Saupe free energy by introducing the effect of *cis* azobenzenes generated from photoisomerization[77, 94]. The model treats the photoactive rod-like *trans* azobenzenes and the non-photoactive mesogens as statistically identical to form the nematic mixture. Assuming first-order light-driven reaction at the single-molecule level, with light polarization considered, the fraction of *cis* azobenzenes is expressed as (see a full derivation in ref.[79])

$$c(Q,\mathbf{n}) = \frac{fI[1 + Q(3\cos^2\beta - 1)]}{3 + I[1 + Q(3\cos^2\beta - 1)]}, \tag{67}$$

where $f$ is the fraction of total photoactive mesogens (*trans* + *cis*), $I$ the dimensionless light intensity, and $\beta$ the angle between the director $\mathbf{n}$ and the light polarization. With this, the Maier-Saupe free energy is modified as[77, 94]

$$W_{\mathrm{lc}}(Q,\mathbf{n}) = N_{\mathrm{n}}kT\big(1 - c(Q,\mathbf{n})\big)\left[g^{-1}(Q)Q - \log Z(Q) - \frac{\big(1 - c(Q,\mathbf{n})\big)J}{2kT}Q^2\right], \tag{68}$$

where the reacted *cis* azobenzenes reduce the total number of mesogens and the effective molecular interaction through a factor $(1 - c)$.

Eqs. 67 and 68 show that light polarization, represented by $\beta$, plays a crucial role in the photomechanical behavior of the LCE. This polarization dependence, which distinguishes photomechanical from thermomechanical actuation, has been investigated extensively in both theory and experiment[35, 79, 95-100]. Furthermore, in Eq. 67, the dimensionless light intensity $I$ is treated as constant, whereas in a real sample of finite thickness, $I$ decays with depth over an attenuation length. This attenuation, coupled to the local photoisomerization, produces a through-thickness gradient in $c$, and hence in $Q$ and in the photomechanical response, which can be captured by solving the coupled attenuation-isomerization equations for $I$ through thickness[98, 101-103].

When the polarization and attenuation of light are neglected, combining Eqs. 67 and 68 shows that the photo-induced reduction in nematic order is equivalent to a rise in temperature[80], consistent with earlier studies on photoactive LCEs[17, 101, 104]. Temperature, however, also governs the entropic elasticity of the polymer network and the *cis*-to-*trans* backward reaction. The resulting three-way thermo-photo-mechanical coupling in photoactive LCEs produces behaviors that do not emerge under either thermal or optical actuation alone[80].

Both the Landau-de Gennes model and the Maier-Saupe model describe a weakly first-order transition, in which the order parameter jumps discontinuously at $T_{\mathrm{ni}}$. Low-molar-mass nematics behave this way, but most LCEs do not: in a crosslinked network, the order parameter is observed to vary almost continuously through the transition, with little or no discontinuity, so that the transition appears second-order or supercritical[105, 106]. This difference is generally attributed to the polymer network, specifically its crosslinking, internal residual stress locked in during synthesis, and coupling between chain stretch and nematic order[18]. In modeling, the second-order transition in LCEs can be accommodated by adding additional energy terms, such as a coupling term $-hQ$, with $h$ representing the stress- or strain-order coupling[1, 18, 107], or physics-based or phenomenological nonideality (as will be discussed in Section 5.3).

Finally, combining the mesogen free energy $W_{\mathrm{lc}}$ with the elastic energy of a monodomain (Section 3) or polydomain (Section 4) LCE gives a total free energy that determines the equilibrium deformation, director, and order parameter simultaneously. Theoretical studies using this total free energy show that the equilibrium order parameter is predominantly governed by $W_{\mathrm{lc}}$ below $T_{\mathrm{ni}}$ and can thus be treated as a control parameter $Q(T)$[78-80]. In contrast, around $T_{\mathrm{ni}}$, the two-way thermomechanical coupling becomes nontrivial: an applied stress raises the effective nematic-isotropic transition temperature, and a change in temperature substantially alters the stress-stretch response[78]. In photomechanics, the combined Maier-Saupe energy, neo-classical energy, and external mechanical load predict a range of behaviors[79, 80, 108]: light-induced contraction, a photomechanical critical point, mesogen rotation and stripe domains driven by mechanical stress and light polarization, temperature-modulated photomechanical actuation, and photo-flexoelectricity.

**5.2 Frank elasticity**

As discussed earlier, the neo-classical monodomain free energy (Eq. 22 or Eq. 51) is non-quasi-convex, such that its direct minimization is ill-posed and must be regularized. A common choice is to introduce additional energy terms from *Frank elasticity* as a regularizer. Frank elasticity originates in the free energy cost of reorienting the director over short distances. In a liquid crystal melt, this Frank energy is generally expressed as[18, 109, 110]

$$W_{\text{Frank}}(\mathbf{n}) = \frac{1}{2}[K_1(\nabla\cdot\mathbf{n})^2 + K_2(\mathbf{n}\cdot\nabla\times\mathbf{n})^2 + K_3(\mathbf{n}\times\nabla\times\mathbf{n})^2], \tag{69}$$

where the three terms correspond to splay, twist, and bend of the director field, and $K_1$, $K_2$, and $K_3$ are material constants of order piconewton[18, 111].

In continuum modeling, $K_1$, $K_2$, and $K_3$ are usually collapsed into a single average material parameter $K_0$, and the simplified Frank energy reduces to[25, 89, 112]

$$W_{\text{Frank}}(\mathbf{n}) \approx \frac{1}{2}K_0\nabla\mathbf{n}:\nabla\mathbf{n} \approx \frac{1}{2}K_0\frac{\partial n_i}{\partial X_K}\frac{\partial n_i}{\partial X_K}, \tag{70}$$

analogous to a surface free energy used in phase field modeling[113, 114].

Taking the Frank constant $K_0 \sim 10^{-12}$ N and the shear modulus of LCE as $\mu \sim 10^4$ Pa, the two material parameters define a length scale $\xi = \sqrt{K_0/\mu} \sim 10^{-8}$ m, i.e., about 10 nm. This material length scale sets the characteristic length of the domain microstructure (such as stripes) in a polydomain LCE: the bulk elastic energy ($\sim\mu$) favors finer textures, while the interfacial cost ($\sim K_0$) between domains favors coarser ones. As a result, $\xi$ represents the length scale over which the director reorients across a domain interface, as well as the thickness of the diffuse interface to regularize the non-quasi-convex energy in numerical models. Since $\xi$ is much smaller than the length scale of the bulk LCE, the effect of Frank elasticity is negligible away from the domain interface.

**5.3 Isotropic-genesis, nematic-genesis, and network nonideality**

The ideal theory predicts a perfect soft-elastic response, with an exactly zero-stress plateau beginning as soon as the sample is stretched and persisting throughout the soft-elastic regime. In real experiments, whether an LCE (monodomain or polydomain) exhibits soft elasticity depends on its specific synthesis and processing. An *isotropic-genesis* LCE has its network polymerized and crosslinked with mesogens in their isotropic phase, either at high temperatures $T > T_{\text{ni}}$ or with solvent. A *nematic-genesis* LCE is polymerized and crosslinked with mesogens in their nematic phase, at low temperatures $T < T_{\text{ni}}$ without a solvent. Experiments have shown that isotropic-genesis LCEs exhibit a semi-soft response[26, 27, 87]: the soft-elastic plateau of the stress-stretch curve begins only after a finite threshold stretch, and the plateau carries a small, rate-dependent, but non-zero stress (e.g., about 0.04 MPa in a thiol-acrylate LCE at 0.01 $s^{-1}$ and 295 K[27]). In contrast, nematic-genesis LCEs exhibit a dramatically higher plateau stress (e.g., about 0.5 MPa in the same thiol-acrylate LCE but after second-stage polymerization in the nematic phase[27]), losing the soft-elastic character.

This discrepancy between the ideal theory and real materials, and between LCEs of different geneses, is attributed to the so-called *nonideality* in real LCEs. One possible physical origin of this nonideality is a residual anisotropy or stress locked into the network by mesogens during the polymerization and crosslinking. An isotropic-genesis LCE retains little such memory of mesogen order, so its response is semi-soft. A nematic-genesis LCE permanently records the director present during the formation of its polymer network, so its response is non-soft[9, 16, 18]. A second possible origin of semi-soft rather than perfectly soft behavior in isotropic-genesis LCEs is the intrinsic heterogeneity of a random network, as in other soft hydrogels and elastomers[115]: because of a wide distribution of chain lengths, network defects (e.g., dangling chains, loops, and entanglements), and mesogen anisotropies, no single soft-elastic deformation is simultaneously optimal for the entire LCE[18].

In continuum theory, this nonideality is usually modeled by introducing phenomenological or physics-based prefactors, or additional free energy terms, into the ideal energy. A widely used form is a nonideality energy[9, 116]

$$W_{\mathrm{ni}} = \frac{\alpha\mu}{2}\mathrm{Tr}[\mathbf{F}(\boldsymbol{\delta} - \mathbf{n}_0 \otimes \mathbf{n}_0)\mathbf{F}^T\mathbf{n} \otimes \mathbf{n}] = \frac{\alpha\mu}{2}[|\mathbf{F}^T\mathbf{n}|^2 - (\mathbf{n}_0 \cdot \mathbf{F}^T\mathbf{n})^2], \tag{71}$$

where $\alpha$ is a dimensionless constant, $\mu$ is the shear modulus of LCE (e.g., $\mu = NkTr^{-1/3}/(1-Q)$ in the neo-classical theory), $\mathbf{n}_0$ is the fixed director imprinted during crosslinking, and $\mathbf{n}$ is the director in the current state. $W_{\mathrm{ni}}$ serves as an energy penalty when the director $\mathbf{n}$ rotates away from the imprinted $\mathbf{n}_0$: it vanishes when $\mathbf{n} = \mathbf{n}_0$ and grows as $\mathbf{n}$ rotates away from $\mathbf{n}_0$. Physically, $\alpha$ can be estimated as $\langle 1/r \rangle - 1/\langle r \rangle$ from a compositional-fluctuation model, in which it represents the spread of chain anisotropy $r$ across the network[9]. When fitted to experimental results, $\alpha$ is on the order of 0.1[25, 117]. More generally, phenomenological fitting parameters and energy terms depending on temperature, stretch, rate, and processing conditions (e.g., pre-stretch) can be introduced beyond Eq. 71 to capture experimental observations across various LCEs[27, 117].

In the literature, Eq. 47 is often presented as the free energy of *nematic-genesis* LCEs, in contrast to Eq. 22 for *isotropic-genesis* LCEs, as the former appears to encode the initial order parameter $Q_0$ and director $\mathbf{n}_0$ in the free energy function. However, this distinction is misleading, since both free energy functions originate from the same biased freely jointed chain model. Consequently, Eq. 47 predicts the same perfect soft elasticity as Eq. 22, leaving no additional energy penalty when the director $\mathbf{n}$ rotates away from the imprinted $\mathbf{n}_0$. Therefore, both Eqs. 22 and 47 describe the ideal monodomain LCE that is more closely approximated by isotropic-genesis LCEs, neglecting the chain nonideality of nematic-genesis LCEs. The apparent difference between the two functions is merely kinematic, arising from their different reference states.

### 5.4 Viscoelasticity

While the theory developed so far has focused on thermodynamic equilibrium behavior, most LCEs are highly viscoelastic[28-33]. In particular, polydomain LCEs show a pronounced rate dependence of the soft-elastic plateau, the subsequent stretch stiffening, the polydomain-to-monodomain transition stretch, and the cyclic loading-unloading hysteresis. The stress level can rise by nearly an order of magnitude over only two decades of stretch rate[28-30, 32, 33], making LCEs among the most dissipative of all stretchable polymer networks.

The strong viscoelasticity of LCEs originates from two intrinsically coupled molecular processes: the ordinary polymer network relaxation induced by chain reptation and intersegment friction[81], and the dissipative deformation-induced reorientation of mesogens[28-30, 118, 119]. Correspondingly, dynamic mechanical analysis (DMA) and stress-relaxation measurements have resolved two well-separated relaxation timescales—a shorter one associated with mesogen rotation and a longer one with network relaxation—each contributing measurably to the rate-dependent mechanical response[28, 118-120]. Theoretically, viscoelastic behavior can be captured by multiplicative decomposition of the deformation gradient into an elastic part and a viscous part. The latter can be modeled by assigning separate viscous elements to mesogen reorientation and network relaxation[30, 31, 34].

The viscoelastic response of an LCE is further jointly regulated by its soft elasticity and temperature. During the stretching of a polydomain LCE, the soft-elastic plateau is associated with mesogen rotation, such that the stress level is strongly rate-dependent and the corresponding relaxation timescale follows time-temperature superposition[24]. Furthermore, the spanning width of the low-rate, quasistatic soft-elastic plateau—the stretch between the undeformed state and the state at which mesogen rotation completes (e.g., $\lambda = r^{1/3}$ in uniaxial stretch of a polydomain LCE, Eq. 59)—is itself governed by the chain-anisotropy ratio $r(T)$ and hence the order parameter $Q(T)$. These two effects together govern various rate-, temperature-, and loading path-dependent behaviors of LCEs, including the cyclic hysteresis, stress relaxation, permanent residual stretch after unloading, and accumulated residual stretch during cyclic loads[32].

The coupling of soft elasticity and viscoelasticity in LCEs facilitates some applications while complicating others. The large cyclic hysteresis of polydomain LCEs makes them attractive for damping, vibration attenuation, and impact protection[52-54]. The dissipation can be further amplified by embedding the LCE in architected structures whose elastic instabilities open a second dissipative channel, yielding energy-absorbing metamaterials and lattices[55-59]. The same viscoelastic dissipation underlies the strong, switchable adhesion of LCE-based adhesives[60-67]. By contrast, in actuation and shape-morphing applications, a monodomain LCE needs to be thermally and/or mechanically cycled repeatedly between shapes, where the strong rate and history dependence becomes a liability.

Despite these recent efforts, several questions remain open in LCE viscoelasticity. First, it remains largely unclear how the individual contributions of different rate-dependent processes govern the LCE behaviors under different loading paths and temperatures. Second, most current viscoelastic models assume an ideal soft-elastic branch, while efforts to combine network nonideality (Section 5.3) with viscoelasticity have only begun[117]. Third, a homogenized description such as the quasi-convex free energy does not resolve the evolving mesogen rotation within domains, which a mesoscale or director-resolved numerical formulation could recover. Finally, high-fidelity, systematic experimental data to construct reliable theoretical models, spanning wide ranges of rate, temperature, and loading path, remain sparse for LCEs.

**6. Conclusion**

We have reviewed the elastic theory of ideal liquid crystal elastomers from the statistics of a single chain to the continuum constitutive model. The statistics of the anisotropic, mesogen-linked Gaussian chains are carried into the thermodynamic free energy through the shape tensor, which contains two physical parameters: the director $\mathbf{n}$ describes the aligned direction of the orientationally ordered mesogens in the nematic phase, and the order parameter $Q$ describes how well the mesogens are aligned. The resulting neo-classical model of a monodomain LCE is non-quasi-convex, admitting a continuous family of stress-free ground states related by director rotation. These ground states give rise to spontaneous deformation, soft elasticity, and coexisting stripe domains, in which the chain-anisotropy ratio $r = (1 + 2Q)/(1 - Q)$ emerges as the key parameter governing the macroscopic response. Quasi-convex relaxation of the same neo-classical energy, obtained by homogenizing over all admissible domain-level director and deformation fields, sorts the macroscopic effective behavior of polydomain LCEs into three mechanical phases. The ideal theory can be further extended to incorporate and describe thermomechanics, photomechanics, Frank elasticity, nonideality, and viscoelasticity. We hope this review serves both as an entry point to LCE mechanics and as a clarification of nuances and ambiguities accumulated in the literature. We further hope that it contributes to the rapid development of LCEs in new theories, experiments, and diverse applications.

**Conflict of interest**

There are no conflicts of interest to declare.

**Acknowledgments**

This work was supported by the National Science Foundation through grants CMMI-2146409 and CMMI-2440758.